\documentstyle[prl,aps,preprint,tighten,floats]{revtex} \begin{document}
\draft

\preprint{IASSNS-HEP-95/102,
Imperial/TP/95-96/14,
hep-th/9512031}
\date{ December 1995}
\title{  Solitonic Strings and
BPS Saturated Dyonic Black Holes
  }
\author{Mirjam Cveti\v c${^1}$\thanks{On
 sabbatic leave from  the University of
Pennsylvania. e-mail: cvetic@sns.ias.edu} and Arkady A. Tseytlin$^2$\thanks{On 
leave from Lebedev Institute,
Moscow. e-mail: tseytlin@ic.ac.uk.}}
\address{$^1$ School of Natural Science\\
 Institute for Advanced
Study, Princeton, NJ 08540, U.S.A.,\\
$^2$ Theoretical Physics Group, Blackett Laboratory \\
Imperial College,   London SW7 2BZ, U.K.\\ }
\maketitle
\begin{abstract}
\baselineskip=12pt
We consider a six-dimensional solitonic string solution described by a
conformal chiral null model with non-trivial $N=4$ superconformal transverse
part. It can be interpreted as a five-dimensional dyonic solitonic string
wound around a compact fifth dimension. The conformal model is regular
with the short-distance (`throat') region equivalent to a WZW theory.
At distances larger than the compactification scale the solitonic string
reduces to a dyonic static  spherically-symmetric black hole of toroidally
compactified heterotic string. The new four-dimensional solution is
parameterised by five charges, saturates the Bogomol'nyi bound and has
nontrivial dilaton-axion field and moduli fields of two-torus.
When acted by combined $T$- and $S$-duality transformations it serves
as a generating solution for {\it all} the static spherically-symmetric
BPS-saturated configurations of the low-energy heterotic string  theory
compactified on six-torus.  Solutions with regular horizons have the global
space-time structure of extreme Reissner-Nordstr\"om black holes
with the non-zero thermodynamic entropy which
depends only on conserved (quantised) charge vectors.
 The {\it independence}
 of the thermodynamic entropy on moduli and axion-dilaton couplings strongly
suggests that it should have a microscopic interpretation as counting
degeneracy of underlying  string configurations.
This interpretation is supported by arguments based on
the corresponding six-dimensional conformal field theory. The expression
for the level of the WZW theory describing the throat region implies a
renormalisation of the string tension by a product of magnetic charges,
thus relating the entropy and the number of oscillations of the
solitonic string in compact directions.
\end{abstract}
\newpage
%\vskip 0.2cm

\baselineskip=14pt
%%%%%%%%%%%%%%%%%%%%%%%%%%%%%%%%%%%%%%%%%%%%%%%%%%%%%
\section {Introduction}
%%%%%%%%%%%%%%%%%%%%%%%%%%%%%%%%%%%%%%%%%
  String theory
is bound  to have important implications
for the physics of four dimensional black holes.
It is likely that certain  fundamental  properties of
 `realistic'  black holes
can be  understood by studying
a special class of
supersymmetric
Bogomol'nyi-Prasad-Sommerfield (BPS) saturated  backgrounds which
for large enough supersymmetry do not receive quantum corrections.
Examples  of  such backgrounds   are provided by pure  electrically
or pure magnetically charged  solutions \cite{gib}
of lowest-order  effective  field
theory  (for a review see \cite{HOR} and references therein).

To embed an effective field theory solution
 into {\it string theory } one is to
find  the corresponding  world-sheet
conformal $\sigma$-model   whose couplings reduce to the given
background fields at scales larger than the compactification and string scales
(see,  e.g.,  \cite{TS} and references therein).
Thus, four-dimensional  effective field theory backgrounds
generically appear to be only
large-distance  approximations  to
higher-dimensional
string solutions.
In particular, all supersymmetric (BPS-saturated) electric
black hole solutions of toroidally compactified
heterotic (or type II superstring) theory \cite{seen}
 correspond to  conformal
chiral null $\sigma$-models   \cite{HRT,TH,BE,CMP,dab}.

The latter   can be interpreted as describing
  higher-dimensional fundamental string
backgrounds, i.e.  external long-range fields produced by  stable
 classical   string sources of  elementary closed
strings, which are in general charged,
oscillate in  one, e.g., left-moving, sector and are
 wound around a compact spatial dimension  \cite{dh,duh,CMP,dab}.
The leading-order solution is
singular at the core\footnote{In fact,
it remains singular to all orders in $\alpha '$ if one
chooses to ignore the source altogether  what, from
the world-sheet $\sigma$-model point of view,
formally seems to be a  legitimate alternative
 \cite{TH,TS}.}
 but
$\alpha'$-corrections most likely provide an
effective smearing of the $\delta $-function source at the quantum
 string  scale $\sqrt{\alpha'}$ \cite{ME}.\footnote{In addition,
  a first-principle conformal field theory interpretation of
solutions with string sources should probably
involve a `thin handle'-type resummation of string
loop expansion \cite{MME}.}
What appears to a distant four-dimensional
 observer as an extreme electric
black hole  has  actually   an internal structure of a higher-dimensional
fundamental string.
This  interpretation
suggests  a natural way of  understanding  the
thermodynamic black-hole entropy
in terms of degeneracy of string configurations
  \cite{sen}, which  give rise to
 black holes  with  the same values of asymptotic charges \cite{CMP,dab}.
 For example,  a
higher-dimensional string  `oscillating' in  a compact internal direction
reduces  to   a  family of black holes
with the same asymptotic charges
but different non-vanishing {\it massive}
 Kaluza-Klein fields which
are invisible   at scales larger than the compactification scale.\footnote{In
order to
 try to reproduce the black hole entropy
as a statistical entropy ($\ln d(N)$)  it  is important that
the oscillating object  should  be a string-like, i.e. having an
 exponentially growing  number $d(N)$ of states
at a given oscillator level $N$;
it is not enough to consider just a Kaluza-Klein theory.}

In order to make this qualitative picture  quantitative, i.e.
to compute the black hole entropy directly from string theory,
one has to address the question of 
string loop and $\alpha'$ corrections.
A nice property of the extreme electric black
 hole (fundamental string) solution is that the dilaton, i.e. the
 effective string coupling,
goes to   zero as one approaches the origin,
so that string loop corrections may be ignored.
This is not true, however, for the $\alpha'$-corrections since
the curvature of the leading-order solution
blows up near the origin.
The expectation \cite{ME} that $\alpha'$-corrections should  smear
the singularity of the fundamental string solution
at scales of order $\sqrt{\alpha'}$
is  in this context closely related
to   the suggestion \cite{sen}
that the thermodynamic black-hole  entropy, which
vanishes when evaluated
at the singular  horizon  ($r=0$)
 of the leading-order effective field theory solution,
should instead be computed at the `stretched'
 horizon  at $r = \sqrt {\alpha'}$ \cite{suss}.
The resulting expression  then matches
 the statistical string entropy \cite{sen,peet}.
Though very plausible, it may be hard to implement
this idea in a  first-principle calculation of the entropy.

Magnetically charged supersymmetric
extreme black holes have very different
properties.
The leading-order solution \cite{gib}  has  a
non-singular  string metric
with the origin   at $r=0$ being transformed
into a `throat' region. Now  $\alpha '$-corrections
can be ignored provided the magnetic charge $P$ is large, i.e.
$P>>\sqrt{\alpha'}$.
Indeed, these configurations have
 the string-theory  representation
\cite{nel,kalor}
  in terms of a higher-dimensional  string soliton,  with magnetic
charge having Kaluza-Klein origin (for a review,  see \cite{CHSR,dukh} and
references
therein). They are
described  by   a regular, source-free  superconformal
field theory \cite{chs}
which reduces in the  throat region to a Wess-Zumino-Witten (WZW) model
supplemented with a linear dilaton.
For that  type of  magnetic  solitons
the dilaton blows up near the origin, and thus
 string loop corrections  cannot  be ignored. This
prevents  one from computing the  black-hole entropy by counting
different solitonic  configurations with the same
four-dimensional large-distance behaviour.

Given the fact that the presence of an electric charge seems
to regularise the short-distance behaviour of the dilaton
while the presence of a magnetic charge leads
to a  regular string metric,
one  may    speculate that to obtain  solutions,
 where both string loop and
$\alpha'$ corrections are under control,  both
  electric and   magnetic charges should be non-vanishing.
Remarkably, this is indeed  what happens in the case
of  four-dimensional
supersymmetric {\it dyonic  black hole solutions}  \cite{CY,CYI}
of leading-order effective field
equations  corresponding to  toroidally
compactified heterotic string
(see  also  \cite{CYII} for a review   and references).
At the string-theory  level they  correspond to
the world-sheet  conformal theory \cite{US}   which
is a hybrid of `electric' chiral null model and
`magnetic'  $N=4$ superconformal model, thus combining  the  best features
of the pure electric and pure magnetic models.
This conformal theory
  describes  a higher ($D \geq 6$) dimensional   string soliton
with all the background fields  regular everywhere (for $r\geq 0$).
The  magnetic charge
plays the role of a short-distance regulator,
 providing an effective shift   $r \to r +P$,
analogous to the shift  $r \to r+ \sqrt {\alpha'}$  expected to happen
in the exact fundamental string solution.

As in the  purely magnetic case, the short-distance region is
a throat  described by a regular
WZW-type conformal field theory,   but now with  a  {\it constant} dilaton.
In fact, the dilaton varies  smoothly  between   constant values
at large and  small distances and its $r=0$ value
is given by
 the ratio of  the magnetic and  electric   charges.
The approximate constancy of the dilaton
ensures that the resulting four-dimensional  dyonic black holes
are black holes indeed; these
solutions have the  global space-time structure of  extreme
Reissner-Nordstr\"om black holes.

As a result, it may be possible  to choose the
 charges   so that
both the  world-sheet $\alpha'$  and  the string loop corrections
{\it   remain small everywhere},
thus  suggesting that  in the case of dyonic
charges one should be able
to reproduce the expression for the black hole entropy
by a  semiclassical computation.
Indeed, the  thermodynamic  entropy  determined by
 the area of the horizon
is now proportional to the product of
the electric and magnetic charges
and thus is  no longer vanishing.
By analogy with the corresponding counting of  degenerate
(fundamental string) states
for purely electric  black 
holes \cite{sen,CMP,dab}
one  may expect   that
the entropy  should now have  an interpretation in terms
of counting of degenerate  solitonic string states \cite{LW}.

To implement this suggestion
it is important to understand
which   solitonic string states
correspond to a given set of asymptotic dyonic black hole charges.
One  should be able to do this by starting directly
with  the underlying  conformal field theory of the dyonic soliton.
As we shall argue below,  for large magnetic charges
the level of the WZW-type conformal field theory
which describes the  horizon (throat) region
is large and thus the counting of states
should be the same as  in flat space
up to a renormalisation of the string tension by magnetic charges, as
anticipated in \cite{LW}.
As a result, one  indeed
reproduces the thermodynamic entropy
by  semiclassical,
string-theory considerations.

The  dyonic model studied in \cite{US}
was   a
six-dimensional  supersymmetric
chiral null model  with curved  transverse part
 which  is a
hybrid of  the
five-dimensional fundamental-string type model (giving rise upon dimensional
reduction along the compact `string' direction
to extreme  black holes with Kaluza-Klein ($Q_1)$  and
 two-form  ($Q_2$)
electric charges)
and an  $N=4$ superconformal  model  (which generalises both
 the Kaluza-Klein monopole ($P_1$)   and the $H$-monopole  ($P_2$) models).
 It has a remarkable  covariance property   with
respect to  $T$-duality in the two compactified dimensions
 and  with respect to  $S$-duality, interchanging the electric
 and magnetic couplings.
To get  a better understanding of  general features
of this class of solitonic conformal models,
in particular,  of their  possible  marginal perturbations,
we  shall generalise  the model of \cite{US}
to include one extra coupling function,  specifying the electric charge  ($q$)
of the solitonic string
(Section II).
The  throat  region of the resulting background is described
by  (an orbifold of) the six-dimensional
$SL(2,R)\times SU(2)$ WZW model
with the level proportional to the product of the two
 magnetic charges $P_1P_2$.
The relation to  the WZW model also implies quantisation
conditions on charges (Section II.B).

In the simplest spherically-symmetric case the
  corresponding
 four-dimensional  dyonic solution
(Section III)
is
parameterised by the
two magnetic $P^{(1)}_1=P_1, \ P^{(2)}_1=P_2$ and the
  two electric  $Q^{(1)}_2=Q_1, \ Q^{(2)}_2=Q_2$ charges
 and  by one new  parameter $q$
specifying  the  electric charges
 $Q^{(1)}_1=-Q^{(2)}_1 =q$   (the upper and lower indices $1,2$
indicate
 the
 Kaluza-Klein and two-form $U(1)$ gauge fields
 and the first and the second compactified
 toroidal coordinates,  respectively).

Like its  $q=0$ limit, corresponding to the  four-parameter
 solution of  \cite{CY,CYI},  the   five-parameter dyonic
solution saturates the Bogomol'nyi bound.
 It  has a non-trivial dilaton, axion
and moduli fields of the compactified two-torus and
 serves, when acted by the $T$- and $S$-duality transformations,
as  a generating solution for  {\it all}
the static spherically-symmetric
BPS-saturated solutions
 of the effective
heterotic string compactified on six-torus. These solutions
are parameterised  by  {\it unconstrained}
28 electric and 28  magnetic  charges (Section IV).
Solutions  with regular event horizons
have the Reissner-Nordstr\"om  global space-time with zero
temperature  and
non-zero thermodynamic entropy.  We derive the
general $T$- and $S$- duality invariant
 expression for  the  entropy, which depends only on conserved
 (quantised) electric and magnetic charges  and is
{\it independent} of  the asymptotic values of the
 dilaton-axion  and moduli fields.
This result supports the expectation  \cite{LW}
that the entropy  is counting the  number
of  string degrees of freedom  which should not
change  under   adiabatic  variations  of couplings of the
theory.

The statistical interpretation of the entropy is discussed
 in Section V by considering the  string-theory
(conformal model) interpretation of the five-parameter solution.
We  present an argument
relating the thermodynamic entropy to the
statistical entropy which  counts  the
degeneracy of the  dyonic solitonic string  configurations
 `oscillating' in a compact direction.
Our approach generalises  the
   suggestion  of \cite{LW}
and  explains the renormalisation of string tension
 by magnetic charges
by direct consideration of the underlying  conformal
model in the horizon (throat) region.

%%%%%%%%%%%%%%%%%%%%%%%%%%%%%%%%%%%%%%%%%%%%%%%%%%%%%
\section {Six dimensional solitonic string  conformal model }
%%%%%%%%%%%%%%%%%%%%%%%%%%%%%%%%%%%%%%%%%%%%%%%%%%%%%%%
The string soliton   we are going to discuss
is described by a  supersymmetric chiral null model
with curved transverse space.
The chiral null models   \cite{TH,TS}  are a  class of  two-dimensional
(2d)
$\sigma$-models  which generalize both plane wave type  and fundamental string
type
models   and  are defined by the following string
Lagrangian:\footnote{We shall use the following notation.
The string world-sheet action
is normalised so that  $I= (\pi \alpha')^{-1} \int d^2\sigma \partial x
{\bar\partial} x 
$ $=(4\pi \alpha')^{-1} \int d^2\sigma \partial_a x \partial^a x$.
The  string  effective action  is
$S_{10}= c \int d^{10} x  \sqrt {-G} e^{-2\Phi} (R + ... )$
and  after reduction to
four dimensions $S_4=(16\pi G_N)^{-1}  \int d^{4} x
 \sqrt {-G'} e^{-2\Phi'} (R' + ... )$,  where the
$D=4$ dilaton $\Phi'(x)$ will be assumed to have trivial asymptotic value, $e^{\Phi'(x\to \infty)}=1$ (so that both string-frame and Einstein-frame 
$D=4$ metrics approach  $\eta_{\mu\nu}$ at infinity), with  $e^{\Phi'_\infty}$ being absorbed in the 
 Newton's
constant $G_N = {1 \over 8} e^{2\Phi'_\infty}\alpha'$.
In Sections III and IV we shall  set
$\alpha ' = 2$ and the compactification 
radii
$R_{n}=\sqrt{\alpha '}=\sqrt 2$. In Section III we shall also assume that 
$e^{\Phi'_\infty}=1$, i.e. that  $G_N = 1/4$.}
\begin{equation}
   L =  F(x)  \partial u \left[{\bar\partial} v +
   K(u,x){\bar\partial}  u  +   2{\cal A}_i(u,x)  {\bar\partial}  x^i \right]
 +   (G_{ij} + B_{ij})(x) \partial x^i {\bar\partial} x^j    +    {\cal R}
\Phi (x)\  .
 \label{lag}
\end{equation}
Here  $u,v$ are `light-cone' coordinates, $x^i$  are
`transverse space' coordinates.  $x^i=(x^s,y^n)$ where $x^s$ ($s=1,2,3)$
are three non-compact  spatial coordinates and  $y^n$ are  toroidally
compactified (Kaluza-Klein) coordinates which
 may also include the   chiral scalar coordinates of the
  internal 16-torus of the heterotic string.
 ${\cal R}\equiv{\textstyle {1\over 4}} \alpha'
\sqrt{ g^{(2)}}  R^{(2)}$ is proportional to the world-sheet curvature.
%$R^{(2)}$.
One can also consider generalisations of this
model by including  $u$-dependence in   functions $\Phi$, $F$, $G_{ij}, B_{ij}$.
Examples
of such conformal  $\sigma$-models (\ref{lag})
with $u$-dependent  $\Phi$  \cite{TH,TS}, \ $F$   \cite{CMP}
and $G_{ij}, B_{ij}$ \cite{TT} were considered in
the literature.

 There exists a renormalisation  scheme
in which (\ref{lag}) is conformal to all
orders in $\alpha'$ provided\   \
 (i) the `transverse'  $\sigma$-model  $(G_{ij} + B_{ij})\partial
x^i {\bar\partial} x^j$
is conformal
 when supplemented with a  dilaton coupling
  $\phi(x)$ \    and \  (ii) the functions
$F^{-1},K,{\cal A}_i, \Phi$ satisfy
the following conditions (as in \cite{US}  we shall  assume that the transverse
theory
  has $N=4$ extended  world-sheet
  supersymmetry so that the conformal invariance
  conditions preserve their one-loop form):
\begin{equation}
- {1\over 2} \nabla^2  F^{-1} +  \partial^i \phi \partial_i F^{-1} =0\  , \ \  
\ \  \
 - {1\over 2} \nabla^2  K    +  \partial^i \phi \partial_i K + \partial_u 
\nabla_i {\cal A}^i
=0 \  ,
\label{ond}
\end{equation}
\begin{equation}
- {1\over 2}  \hat  \nabla_i {\cal F}^{ij} + \partial_i \phi {\cal F}^{ij}
=0  \  ,  \ \ \ i.e.  \ \  \
\nabla_i (e^{-2\phi}{\cal F}^{ij}) -  {1\over 2} e^{-2\phi}{\cal F}_{ik} H^{ikj}=0\ , 
\label{cond}
\end{equation}
with
\begin{equation}
  \hat \nabla\equiv  \nabla (\hat \Gamma)\ , \  \ \ \hat \Gamma^i_{jk}
=  \Gamma^i_{jk} + {1\over 2} H^i_{\ jk} \ ,  \ \ \ \
 {\cal F}_{ij} \equiv  \partial_i {\cal A}_j -
\partial_j {\cal A}_i\ ,
\  \  \  \ \ \
 \Phi =  \phi  + {1\over 2} \ln F\   ,
\label{few}
\end{equation}
where $\nabla_i$ is covariant derivative defined  with respect to the
transverse metric $G_{ij}$ and $H_{ijk} = 3 \partial_{[i} B_{jk]}$.
The Maxwell-type  equation for ${\cal F}_{ij}$ is the conformal invariance
condition in the $(ux^i)$-direction.
The linear equations for $K$ and ${\cal A}_i$ can
 be viewed as marginality conditions on  the  corresponding
 `perturbations'
of the conformal model specified by $F, G_{ij}, B_{ij}, \phi$.

Given a $u$-independent solution of the above equations
one can construct its
 $u$-dependent generalisation, e.g.,
by replacing the constant parameters in $K$ and ${\cal A}_i$
by functions  of $u$, i.e.
 $K(x;c) \to K(x; c(u)),$
 and $ {\cal A}_i (x;q)  \to  h_i(u) + {\cal A}_i(x;q(u))$.
If $\partial_u \nabla_i {\cal A}^i \not=0$ the function
$h_i$ will be related
to $K$ by the second equation in  (\ref{ond}).
In the case of the fundamental string solution  with $u=x_9-t,\ v=x_9+t,$
and $G_{ij}+B_{ij}=\delta_{ij}$
this corresponds to adding \cite{garf,TH,CMP,dab}
traveling waves of momentum along the string as well as
arbitrary
left-moving oscillations both in compact  $y^n$ (charge)  directions
and  non-compact spatial directions $x^s$.
Such solutions are  in  correspondence with BPS-saturated states
of the heterotic string spectrum
at the vacuum level   in
the right-moving  sector  (right-moving oscillator number $N_R={1\over 2}$)
and an arbitrary level in the left-moving sector
(arbitrary left-moving oscillator number $N_L$) \cite{CMP,dab}.

Let us now specialise to the particular case  of six-dimensional $\sigma$-model
(\ref{lag}) of the type:
\begin{equation}
 L = F(x)  \partial u \bigg({\bar\partial} v +  K(x) {\bar\partial} u
        + 2A(x)[{\bar\partial} x^4 + a_s (x)  {\bar\partial} x^s]\bigg)  + 
{1\over 2} {\cal R} \ln F(x)
+ L_{\bot}\ ,
\label{lagr}
\end{equation}
$$
L_{\bot}=  f(x)k(x)  \big[ \partial x^4 + a_s (x) \partial x^s\big] \big[
{\bar\partial} x^4  + a_s (x) {\bar\partial} x^s\big]
    + f(x)  k^{-1} (x) \partial x^s {\bar\partial} x^s
$$
\begin{equation}
 +\   b_s (x) (\partial x^4 {\bar\partial} x^s - {\bar\partial} x^4 \partial 
x^s)
  +   {\cal R}   \phi(x) \  , 
 \label{latra}
\end{equation}
where  $x^s=(x^1,x^2,x^3)$ are three non-compact spatial dimensions,
while  $x^4$ and $u$ will  be compact coordinates. We shall assume
that  all the fields  depend only on  $x^s$  and $f,k,a_s,b_s,\phi$
are subject to\footnote{More
generally,  one may consider a similar  six-dimensional model  with the
functions depending on all four transverse coordinates.
Special cases will be the six-dimensional fundamental string ($F^{-1}= 1 +
Q/x^2$, $K=f=k=1$) \cite{duh},
its $S$-dual solitonic
string  solution
%of type II theory compactified on K3
 \cite{dukh,senn,HAS} ($F=K=1, A=0, \ f= 1 + P/x^2, k=1$)
which  also  corresponds to a  six-dimensional reduction
 of the five-brane solution  \cite{chs,DL}
of  the ten-dimensional theory
 and the dyonic
six-dimensional string  \cite{duf} ($F^{-1}= 1 + Q/x^2,
\ f= 1 + P/x^2,\  k=K=1$).}
\begin{equation}
   \partial_s\partial ^s f =0  ,  \ \ \ \ \ \ \ \ \ \ \ \    
\partial_s\partial^s
   k^{-1} =0,
\label{hpp}
\end{equation}
\begin{equation}
 \partial_{p} b_q - \partial_q b_{p}
 = -\epsilon_{pqs} \partial^s   f   , \ \ \
\partial_{p} a_q - \partial_q a_{p}
 =- \epsilon_{pqs} \partial^s   k^{-1} , \ \  \ \phi= {1\over 2} \ln f ,
\label{hypp}
\end{equation}
where $p,q,s=1,2,3$.

In (\ref{lagr}) we have made a special choice of the field
${\cal A}_i$:
\begin{equation}
{\cal A}_s=A a_s, \ \ \  \ \ {\cal A}_4 \equiv A,
\label{gf}
\end{equation}
which makes
the model covariant under the duality transformation in the $x^4$-direction.
The 2d duality transformation  $x^4\to \tilde x^4 $
can be  performed by  gauging the shifts in $x^4$,
adding $ B\partial \tilde x^4 - \bar B \partial\tilde x^4$,
gauge-fixing $x^4=0$ and integrating out $B$ and $\bar B$.
One finds that indeed  this duality transformation
 maps the model (\ref{lagr}) into itself
with\footnote{Note that under the duality in $u$-direction
$F \to K^{-1} ,    \   K \to F^{-1} ,\   \Phi (F) \to \Phi(K^{-1})$ while other
functions
 remain unchanged.}
\begin{equation}
f \to k^{-1}\ , \ \  \   k \to f^{-1}\ ,
\ \ \  a_s \to  b_s\ , \ \  \ b_s \to
a_s\  , \ \ \ A\to  (fk)^{-1} A  \ ,
\label{dua}
\end{equation}
where we have assumed that  $a_{[p} b_{q]}=0$.
The  choice  of ${\cal A}_i$ (\ref{gf})
also leads to the
absence of a  Taub-NUT term in the metric of the  corresponding
four-dimensional spherically-symmetric background.

Let us note that
 case of six
 target space  dimensions
 is special in that here  the duality transformation
applied to  a  rank-two antisymmetric tensor  gives again a rank-two tensor
and thus
can be represented as a formal  map of  one string $\sigma$-model into another.
 In contrast to $T$-duality, however, this
transformation cannot be realised
directly at the 
world-sheet level.\footnote{This transformation
maps one solution (with vanishing vector fields) of the toroidally compactified 
to  six dimensions
heterotic  (or type II)  string into another. 
In the context of string-string duality  between heterotic string 
on  four-torus  and type II string  on $K3$ surface
 \cite{HTI,WITTENII}
this transformation  should be supplemented by  identification of the  vector fields in  the
Neveu-Schwarz--Neveu-Schwarz 
 sector  of the heterotic string   with the 
 the vector  fields in the Ramond-Ramond   sector of 
 the type II string.}
As was pointed out in \cite{US},  the model
(\ref{lagr}),(\ref{latra})  with  $A=0$
has  a remarkable covariance  property
 under  the  six-dimensional  $S$-duality ($ G \to  e^{-2\Phi} G$,  \
$ dB \to  e^{-2\Phi} \ast dB, $
\ $\Phi \to -\Phi$):  when formally applied to the background
fields  of the $\sigma$-model this transformation simply interchanges the 
functions
$F$ and $ f$.
 When $A\not=0$
the above six-dimensional model is  still covariant under  the $T$-duality
  in $x_4$ and $u$ directions.  However,  it is no longer covariant
 under the  $S$-duality.
 The reason is that for $A\not=0$  the components $H_{u4s}$ and $H_{upq}$
 of the torsion  become non-vanishing but
 under  the duality  they are transformed into
$H'_{v4s}$ and $H'_{vpq}$.
That means that the $S$-duality induces the torsion  terms
$\sim \partial v\partial x^4$ and $\sim \partial  v\partial  x^s$  in the
$\sigma$-model action.
Though the resulting background
will again  represent  a leading-order  solution of the string effective
equations,
now it is  not  clear whether it will
remain an exact  solution to all orders in $\alpha'$.

The conditions (\ref{ond}) on
$F,K$ and on ${\cal A}_i$  (i.e. on $A$)
can be put into the form
\begin{equation}
\partial_s\partial ^s  F^{-1} =0 \  , \ \ \ \  \ \partial_s\partial^s  K =0 \ ,
   \label{liin}
\end{equation}
\begin{equation}
 \partial_s (k^{-1}f^{-1} \partial^s A )
+ k^{-1}\partial_s k \partial^s (k^{-1}f^{-1}) A=0
\ , \ \  i.e. \ \ \ 
\partial _s [k^{-3}f^{-1} \partial^s (kA) ]=0 \ .
\label{llel}
\end{equation}
In deriving (\ref{llel})
 from (\ref{cond}) we have used that $a_s,b_s$ satisfy
(\ref{hypp}) and that $f$ and $k^{-1}$ are harmonic.
The model is thus parameterised by four harmonic functions $F^{-1},K,
f,k^{-1}$  and the  function $A$ satisfying (\ref{llel}). The functions $a_s$
and $b_s$  are then
determined by $f,k$ according to (\ref{hypp}).

 The model discussed above is a generalisation of the one
introduced
in \cite{US} where the fifth function $A$ was turned off. Let us
emphasize that   unlike $F,K$ and $f,k$ terms,  which
 coexist in (\ref{lagr})  without influencing the  equations  of each
 other,   the  introduction of the new  coupling $A$
leads to the  equation  (\ref{llel})  which {\it  depends} on the
couplings
$f,k$  of the transverse part of  the $\sigma$-model (\ref{lagr}).

%%%%%%%%%%%%%%%%%%%%%%%%%%%%%%%%%%%%%%5
\subsection{Solution of conformal invariance conditions}
%%%%%%%%%%%%%%%%%%%%%%%%%%%%%%%%%%%%%%%%%%%%%%%%%

While $F^{-1}, K, f, k^{-1}$ are  {\it independent} harmonic functions,
$A$ is specified by   (\ref{llel})  which    has the following  solution:
 \begin{equation}
 A= q_1 k^{-1}  + q_2 f^2 k    , \  \ \ \  q_{1, 2} ={\rm const}  .  
\label{spec}
\end{equation}
This is the general solution in the case of one-center spherically-symmetric
harmonic functions $f,k^{-1}$.
For more general $f,k^{-1}$ , e.g., multi-center harmonic functions,
the solution of the linear equation (\ref{llel})
 (which is equivalent to the  scalar Laplace equation for $kA$
in curved three-dimensional space with
conformally-flat metric $ds^2_3= \rho^2 (x) dx^sdx^s, \ \rho = k^{-3}f^{-1}$)
 will look more complicated.

 If we
further assume the asymptotic flatness conditions on the functions, i.e. that
$k\to 1, \ f\to 1, \ A\to 0$  for
$r^2\equiv x^sx_s\to \infty$,   then  (\ref{spec}) becomes
\begin{equation}
  A=  q_0  k (k^{-2} -f^2)    , \ \ \ \  \ \  q_0=q_1=-q_2   .
\label{spee}
\end{equation}
Note that under the $T$-duality transformation  (\ref{dua})
the $(q_1,q_2)$ solution  for $A$ is mapped into the $(q_2,q_1)$ solution,
i.e. $q_0$ in (\ref{spee}) changes sign.

In the special case of one-center harmonic functions
we get the following explicit form of the solution:\footnote{The relation to 
the
notation used in \cite{US}  is: $Q_1={{\bf Q}^{(1)}_2},
\ Q_2={{\bf Q}^{(2)}_2}, \ P_1={{\bf P}^{(1)}_1},
\ P_2={{\bf P}^{(2)}_1}$.}
\begin{equation}
F^{-1}  =  1+{Q_2\over r}\  ,\ \ \
  K=1+{Q_1\over r}\ ,  \ \ \
f = 1+{P_2\over r}\ ,\ \ \
k^{-1} = 1+{P_1\over r}\ ,
\label{FKfk}
\end{equation}
\begin{equation}
A= {q\over r} \cdot { r +  {1\over 2} (P_1+P_2) \over r + P_1} \  , \
%\ \ \ \ q\equiv 2q_0 (P_1-P_2),  \ \  P_+\equiv  {1\over 2} (P_1 +P_2) \ ,
\label{aaa}
\end{equation}
\begin{equation}
 a_s dx^s= P_1 (1-\cos \theta) d\varphi\ , \ \ \ \  \ \  b_sdx^s = P_2 (1-\cos
\theta)d\varphi\ ,
\end{equation}
\begin{equation}
e^{2\Phi} = F e^{2\phi}= {{ r + P_2}\over{r + Q_2}} \ ,
\label{diit}
\end{equation}
where the parameter $ q$ is  related to $q_0$ of (\ref{spee})  as $q\equiv 2q_0
(P_1-P_2)$.  Note that the expression for $A$ in terms of $q$   is valid also
for  $P_1=P_2$, i.e. for $fk=1$, when it becomes just the  harmonic function
$A= {q/ r}$.
The  one-center solution is thus  specified by the five  parameters
$P_1,P_2$, $Q_1, Q_2$ and $q$.

Let us note that  for positive $P_2$ and
$Q_2$  the string  dilaton  (\ref{diit})
  is regular and is constant both at
 large  $r\to\infty$  and  small   $r\to 0$ distances.
Thus one  can, in principle,
 make the  effective string coupling
  small everywhere by choosing  $e^{\Phi_\infty} < 1 $ and 
 $P_2 \sim Q_2$.
%%%%%%%%%%%%%%%%%%%%%%%%%%%%%%%%%%%%%%%%%%%%%%%%

%%%%%%%%%%%%%%%%%%%%%%%%%%%%%%%%%%%%%%%%%%%%%%%%%%%%
\subsection{Throat region }
%%%%%%%%%%%%%%%%%%%%%%%%%%%%%%%%%%%%%%%%%%%

An important property  of the
six-dimensional $\sigma$-model  (\ref{lagr}),(\ref{latra}), 
(\ref{FKfk})-(\ref{diit})
is that in contrast to the six-dimensional  chiral null model with flat
transverse part
 ($f=k=1$)  which is singular
at $r=0$  (and describes a fundamental string type configuration),
in the case of  non-trivial transverse part  with
 non-vanishing parameters  $P_1>0$ and $P_2>0$
 the singularity at the core $r=0$
disappears.  It gets replaced by a `throat'  or `semi-wormhole' region
\cite{chs}.

In the  throat
region $r\to 0$ the  Lagrangian (\ref{lagr}) (with  the functions given by
(\ref{FKfk})-(\ref{diit}) and  $P_1,P_2,Q_1,Q_2$ all positive) takes the form
$$
 L_{r\to 0} =  P_1
 %^{-1} 
 P_2  \partial z {\bar\partial} z +  e^{-z}   \partial u 
{\bar\partial} v +
  Q_1 Q_2^{-1}  \partial u {\bar\partial} u
        + 2 Q_2^{-1} q \partial u [{\bar\partial} y_1 +
         P_1 (1-\cos \theta) {\bar\partial} \varphi]
$$
$$ + \  P_1
^{-1}
P_2 [\partial y_1 + P_1 (1-\cos \theta) \partial \varphi][{\bar\partial} y_1 + 
P_1 (1-\cos
\theta) {\bar\partial} \varphi]       +   P_1
%^{-1}
P_2(\partial\theta{\bar\partial}\theta+\sin^2\theta\partial\varphi
{\bar\partial} \varphi)
$$
\begin{equation}
 +\  P_2 (1-\cos \theta) (\partial y_1 {\bar\partial} \varphi - {\bar\partial} 
y_1 \partial \varphi)
   , \ \ \ \ \ \  z \equiv - \ln {r \over Q_2} \to \infty \ .
\label{thro}
\end{equation}
In the case of  $q=0$  discussed in \cite{US}
 we get  a  regular conformal   model which, up to a
 factorisation over a discrete subgroup, is
the WZW theory based on  a direct product of the $SL(2,R)$
and $SU(2)$ groups.\footnote{In the special  case 
of $P_1=P_2$, $Q_1=Q_2$, $q=0$  similar 
throat region  model 
was  considered previously  in
\cite{LS}.}
An  important feature  is that,
in contrast to, e.g.,   the five-brane model \cite{chs},
here the dilaton (\ref{diit})
is {\it constant}
in the throat region, i.e. the string coupling is not blowing up
and thus the  solution can be trusted everywhere.\footnote{One
manifestation of  the regularity of the
 dilaton in this  model
is that the
central charge of this conformal field theory,  which has a
free-theory value in the supersymmetric $\sigma$-model case,
can be easily computed either
in  $r=\infty$ or in $r=0$ regions  \cite{US}.}

For a non-zero $q$  it looks as if
the Lagrangian (\ref{thro}) describes
a globally non-trivial
  `mixture' of the   $SL(2,R)$ and $SU(2)$
theories. However, the  central charge  retains its
free-theory value, i.e.  the dilaton $\Phi$ is still constant at $r=0$,
and
it is easy to see  that (\ref{thro})
can, in fact,  be put in the same form as in  the $q=0$  case \cite{US}
by redefining the  coordinates.
Changing the notation for coordinates  to
$u=y_2, \ v=2t, \ x_4=y_1$
where
$y_1$ and $y_2$  will be assumed to be circular  coordinates
with   periods  $2\pi R_1$ and $2\pi R_2$
we get (up to a total-derivative term
$\propto  q(\partial y_2{\bar\partial} y_1 -{\bar\partial} y_1 \partial y_2)$)
\begin{equation}
 L_{r \to 0}  =   \left(p \partial z {\bar\partial} z +  p' \partial \tilde 
y_2  {\bar\partial}
 \tilde y_2
 +  2  e^{-z }   \partial \tilde y_2 {\bar\partial} \tilde  t  \right)
\label{ttr}
\end{equation}
$$ +   \   p \left( \partial \tilde y_1 {\bar\partial} \tilde y_1     + 
\partial \varphi {\bar\partial}
\varphi +
\partial \theta {\bar\partial} \theta  - 2 \cos \theta  \partial  \tilde y_1 
{\bar\partial} \varphi \right)  \ .
$$
Here
 \begin{equation}  \tilde y_1 = { {P}_1^{-1} } y_1  +  {q P_2^{-1} } \tilde 
y_2  +
 \varphi ,
\ \  \tilde
 y_2 = { {Q} _2^{-1} } y_2, \ \  \tilde t = Q_2 t ,  \ \  p= {P} _1 {P} _2,  \ \
 p'={Q} _1 { Q} _2 -  q^2  {P_1 P_2^{-1} }.
\label{not}
\end{equation}
The role of $q$ is thus to mix the two compact coordinates $y_1$ and $y_2$,
i.e. in the throat region  $q$  plays a role of a  modulus, which turns on the
off-diagonal components of the metric of the
 two-torus corresponding to $y_1,y_2$.

The  throat region model (\ref{ttr})
is  thus equivalent to a direct product
of   the   $SL(2,R)$ and $SU(2)$ WZW  theories
 (corresponding to the  terms  in each of the
  parentheses in (\ref{ttr})) divided  by discrete subgroups.
The levels   of the  $SL(2,R)$ and $SU(2)$   models
 are
 both equal to
\begin{equation}
 \kappa= {4\over \alpha'}p = {4\over \alpha '}  P_1P_2 \ .
\label{lev}
\end{equation}
Since the level of  $SU(2)$  must be integer, we get
the quantisation condition $P_1P_2 = {\textstyle{1\over 4}} \alpha' \kappa  $.

When $q=0$
one can follow \cite{gps,nel}
 and construct an orbifold
of $SU(2)$ by identifying the coordinate
$\tilde y_1$,  which in the standard $SU(2)$ WZW model must be 
$4\pi$-periodic, in
the following way:
${\tilde y}_1\equiv {\tilde y}_1+4\pi/m$, where $m$ is an
 integer.
This is possible provided\footnote{The quantisation of $P_1$  can  also be 
understood
 as  a  consequence  of  the requirement of regularity of the
metric ($\sim [dy_1 + P_1(1-\cos \theta )d\varphi]^2 + ...$)
of  the full six-dimensional  model (\ref{lagr}):
  to avoid the  Taub-NUT
singularity
one should be able to identify $y_1$
with  period $4\pi P_1$,  which is possible if  $2P_1/R_1 = m$.
By $T$-duality, the same constraint  should also apply  to
$P_2$, i.e.  $2P_2= n  {\tilde R_1}= n\alpha '/R_1$. }
\begin{equation}
 2P_1= m R_1 \ .
\label{qua}
\end{equation}
Then the  modular invariance of the orbifold $SU(2)_\kappa/Z_m$ demands
 \cite{gps}
that $\kappa= n m $  where $n$ is an  integer.
Since
here
the level $\kappa$ of $SU(2)$  is itself proportional to the product
of the two magnetic charges,
we also get the following quantisation  condition for $P_2$: \
\begin{equation}
 2P_2={{ n \alpha '}\over{ R_1}}\ .
\label{quas}
\end{equation}
For $q\not=0$ we
  demand that  the coordinate $\tilde y_1$
should still have the same  period $4\pi/m$ and
  get  an    additional  condition ${2 P_2 Q_2}    =  l m qR_2$,
where $l$ is an integer, i.e.
\begin{equation}
{Q_2 \over q} = {lm\over n } = {lP_1\over  P_2}   \ .
\label{coo}
\end{equation}
$T$-duality in $y_2$-direction,  which implies $Q_1 \to Q_2, \ 
R_2 \to
\alpha '/R_2$,
yields
\begin{equation}
{Q_1 \over q} = { {l'm}\over n}
  = {{l'P_1}\over P_2}
\ .
\label{cod}
\end{equation}
 Thus
 the consideration of the
throat region leads to  the   relations which
 {\it mix} the quantisation conditions on  $q$, $Q_n$ and $P_n$.

%%%%%%%%%%%%%%%%%%%%%%%%%%%%%%%%%%%%%%%%%%%%%%%%%%%%%%%%%
\section{Four dimensional dyonic  black holes }
%%%%%%%%%%%%%%%%%%%%%%%%%%%%%%%%%%%%%%%%%%%%%%%%%%%%%%%%%%

The  six-dimensional $\sigma$-model of the previous Section can be interpreted 
as
 describing a  dyonic  five-dimensional  ($u,v,x_s)$
solitonic string solution.
  The  string has both  electric ($q$) charge
 (with the  gauge field one-form
 $ Adu$) and magnetic ($P_1$) charge, both
resulting  from the  couplings  to the  compact Kaluza-Klein
$x_4\equiv y_1$ direction.  There is also
 a momentum ($Q_1$) along the string
and  a  special perturbation ($P_2$)  in  curved non-compact spatial
directions $x_s$.
 The  new coupling
$A$    preserves the  same amount of the
 world-sheet supersymmetry of
the  
%RNS  
$\sigma$-model string action, and thus the same amount of  the  space-time
supersymmetry of the target space  background,
as  in the case  of  $A=0$  \cite{US}.

Like the fundamental string  solution  \cite{dh,duh}
$(f=k=1)$ can be viewed
as a field  of  stable elementary winding string mode
in flat background,
this solution may be interpreted as
 representing  
%  excitations of 
a particular state of a  dyonic  string soliton. This interpretation is consistent
  with  the linear form of the  equations for $F^{-1}, K, A$
(\ref{liin}),(\ref{llel})
which
 can be viewed as  conditions
of marginality for (exact) perturbations of
the six-dimensional model defined as
a direct product of the  trivial chiral null (`electric')  part, specified
by
$u,v$, and  non-trivial transverse  (`magnetic') part, specified by $x_s,x_4$.

At the same time, this  model has also
 a  four-dimensional  dyonic black hole
interpretation.
 We shall now derive the explicit expressions
for the corresponding canonical four-dimensional fields.

%%%%%%%%%%%%%%%%%%%
\subsection{Dimensional reduction}
%%%%%%%%%%%%%%%%%%%%%%%%%%%%%%%%%%%%%%%%%%%%%%%%%%%%%%%%%%%%%
Following \cite{TH}, we  set $u=y_2, \ v=2t, \ x_4=y_1$
where $y_1,y_2$ are two  compact   toroidal coordinates.\footnote{In
the case
of the fundamental string interpretation
 the direction  of the string winding is
the `boosted'
compact coordinate   $y'_2=y_2 + t$, i.e.  $u=y_2'- t, \ v =y_2'+ t$.
Then, e.g., for the one-center  solution one gets    $K=Q_1/r$
instead of $K=1 + Q_1/r$  used  here.}
One can rewrite the $\sigma$-model (\ref{lagr}), i.e.
$L= (G_{MN} +B_{MN} )\partial x^M {\bar\partial} x^N +  {\cal R} \Phi $,
 in the form\footnote{In the
purely `electric' case of $f=k=1$, i.e. the case of the
chiral null model
with flat transverse part,
 similar dimensional
reduction was discussed in \cite{BE}.
There, an additional term
 $2F{\cal A}_s \partial u{\bar\partial} x^s$
  was  also included,
leading to   non-static,
e.g., Taub-NUT or rotating,    four-dimensional space-time metric.
}
$$
L =   (G'_{\mu\nu} + B'_{\mu\nu}) (x) \partial x^\mu {\bar\partial} x^\nu
+
 (G_{mn} + B_{mn})(x) [ \partial y^m   + A^{(1)m}_{\mu}(x)  \partial x^\mu]
  [ {\bar\partial} y^n   + A^{(1)n}_{\nu}(x)  {\bar\partial} x^\nu]
$$
\begin{equation}
+ \ A^{(2)}_{n\mu}(x)  (\partial y^n {\bar\partial} x^\mu - {\bar\partial} y^n 
\partial x^\mu)
+  {\cal R} \Phi'(x)  ,
\label{kkk}
\end{equation}
where $x^\mu= (t, x^s),$ $  s=1,2,3, $
 $ \ n,m=1,2$.
The four-dimensional string-frame space-time metric $G_{\mu\nu}'$, the
two form-field $B_{\mu\nu}'$ and the dilaton $\Phi'$,
which includes  the shift  resulting from
 `integrating out'  $y^n$,  as well as the canonical
vector potentials $A^{(1)n}_\mu$ and  $A^{(2)}_{n\,\mu}$ of the respective
Kaluza-Klein  and two-form $U(1)$ gauge fields
are related to the fields of the six-dimensional
$\sigma$-model (\ref{lagr}) in  the following way:
\begin{equation}
G'_{\mu\nu} = G_{\mu\nu} - G_{mn} A^{(1)m}_{\mu} A^{(1)n}_{\nu}, \ \ \
B'_{\mu\nu} = B_{\mu\nu} -  B_{mn}A^{(1)m}_{\mu} A^{(1)n}_{\nu},
\label{kkkk}
\end{equation}
\begin{equation}
A^{(1)n}_{\mu} =  G^{nm} G_{m\mu} \ , \ \ \
A^{(2)}_{n\mu } =  B_{n\mu} - B_{nm} A^{(1)m}_\mu , \ \ \
\Phi' = \Phi  -{1\over 4} \Delta, \ \ \  \Delta\equiv \det G_{mn} .
\end{equation}
The four-dimensional Einstein-frame metric
is
\begin{equation}
g_{\mu\nu} = e^{-2\Phi'} G'_{\mu\nu} \  ,
\label{EE}
\end{equation}
 and
the gauge-invariant  torsion  can be
written as \cite{MS}:
\begin{equation}
H'_{\mu\nu\lambda} = H_{\mu\nu\lambda} - (A^{(1)n}_{\mu} H_{n\nu\lambda}  -
A^{(1)m}_{\mu} A^{(1)n}_{\nu}  H_{mn\lambda} + {\rm cycl.\  perms.}),
\label{tor}
\end{equation}
where $H_{MNK}$ is the field strength of the antisymmetric tensor $B_{MN}$ in
(\ref{lagr}).
$H'_{\mu\nu\lambda}$ is related to   the four-dimensional axion  $\Psi$ by
 \begin{equation}
{H'}^{\mu\nu\lambda}\equiv  {e^{4\Phi'} \over \sqrt{
-g}}   \epsilon^{\mu\nu\lambda\rho} \partial_\rho \Psi\ ,
\label{axx}
\end{equation}
where the indices are raised  using $g_{\mu\nu }$ and  $g={\rm det}g_{\mu\nu}$.

%%%%%%%%%%%%%%%%%%%
\subsection{Four-dimensional background}
%%%%%%%%%%%%

We shall  now  express  the four-dimensional
fields in terms of  harmonic functions $F^{-1}$, $K,$ $f$, $k^{-1}$, the
function $A$ and the functions $a_s,b_s$.

The  Einstein-frame metric  is found to be  of the following form:
\begin{equation}
ds^2_E =g_{\mu\nu}dx^{\mu}dx^{\nu}=- \lambda (r)  dt^2 + \lambda^{-1} (r)  
(dr^2 + r^2
d\Omega^2) \ .
\label{eee}
\end{equation}
The structure of the space-time metric is that of an {\it extreme} static 
spherically
symmetric configuration. This  indicates  that this background
corresponds to  a  BPS-saturated state.

The  metric  function $\lambda$, the dilaton $\Phi'$, and the moduli  $G_{mn},
B_{mn}$ of the
two-torus  are given by
\begin{equation}
\lambda=Fk\Delta^{-1/2}, \ \ \ e^{2\Phi'}=F f \Delta^{-1/2}, \ \ \ \Delta=  F 
K f k -
A^2 F^2,
\end{equation}
\begin{equation}
 G_{11}=fk,  \ \ \ G_{22}=FK, \ \ \  G_{12}=-B_{12}= AF.
\label{modd}
\end{equation}
The  description in terms of the four-dimensional fields  is valid
in the spatial region,  where  $F,K,f,k$
and
the volume of the two-torus $\Delta$  are all  positive.
 The  constraint $\Delta > 0 $ implies
a constraint on the  function $A$: \  $F^{-1} Kfk > A^2 $.

The four  four-dimensional $U(1)$  Kaluza-Klein and two-form
 gauge fields  have the following components
\begin{equation}
A^{(1)1}_{\mu}=  (-AF^2\Delta^{-1},\  a_s), \ \ \ \
A^{(1)2}_{\mu}= (Ffk\Delta^{-1}, \ 0) ,
\label{fie}
\end{equation}
$$
A^{(2)}_{1\mu  } = ( AF^2fk\Delta^{-1}, \ b_s)  , \ \ \ \
A^{(2)}_{2\mu } = ( F^2Kfk\Delta^{-1} , \ 0) .
$$
The axion $\Psi$ is determined by:
\begin{equation}
\partial_s\Psi=Af^{-2}k^{-1} \partial _s (fk ) \ .
\end{equation}
%%%%%%%%%%
\subsection{One-center four dimensional solution}
%%%%%%%%%%%%
In the case of spherically-symmetric one-center harmonic functions
$F^{-1},K,f,k^{-1}$
 the explicit solution  (\ref{FKfk})-(\ref{diit}) yields
a class of static  spherically-symmetric  four-dimensional
backgrounds   specified by  the  five parameters $P_1,P_2$, $Q_1,Q_2$
and $q$.\footnote{The five-parameter
  extreme (BPS-saturated)  solution
{ as well as} non-extreme  solutions of the effective
heterotic  string  action
compactified on  six-torus were obtained  independently in \cite{CYIII} by
performing a subset of $O(8,24)$  symmetry transformations of  the
three-dimensional effective action  on   the Schwarzschild black hole
background.
They   serve as generating solutions for all the  static spherically-symmetric
configurations of the  heterotic string theory
 compactified on six-torus.
 The BPS-saturated solution obtained in \cite{CYIII}
is  related to the one described in this Section through a subset of
$ SO(2)\times SO(2)\subset
O(2,2)$ ($T$-duality)  and  $SO(2)\subset SL(2,R)$ ($S$-duality)
transformations.}
These parameters determine the magnetic  $P^{(1,2)}_m$  and electric
$Q^{(1,2)}_m$ charges of the  corresponding Kaluza-Klein  $A^{(1)\,m}_\nu$ and
two-form  $A^{(2)}_{m\nu}$ gauge fields, i.e. $A^{(i)}_{m\, t}\to  
-Q^{(i)}_m/r$
as $r^2\equiv x_sx^s\to \infty$, and $A^{(i)}_{m\,\phi}=(1-\cos\theta )
P^{(i)}_m$.  As follows from
(\ref{FKfk})-(\ref{diit}) and the expressions for
 the gauge fields (\ref{fie}),
the physical charges are related to these  five parameters in the following
way:\footnote{In this Section we assume that 
$\alpha ' = 2 $, $e^{\Phi'_\infty}=1$,  the Newton's constant $
G_N = {\textstyle {1\over 8} } \alpha ' e^{\Phi'_\infty} = {1\over 4}$ and the compactification 
radii
$R_{1}=R_2=\sqrt{\alpha '}=\sqrt 2$.}
 $$(Q^{({1})}_1, P^{(1)}_1)=(q,P_1), \ \  \ \
(Q^{({1})}_2, P^{(1)}_2)=(Q_1,0),
$$\begin{equation}
(Q^{({2})}_1, P^{(2)}_1)=(-q,P_2), \ \  \ \
(Q^{({2})}_2, P^{(2)}_2)=(Q_2,0) .
\label{chr}
\end{equation}
Note that  the magnetic charges
arise from the transverse part of the $\sigma$-model (\ref{lagr}) and  all
the electric charges arise from the chiral
null part of (\ref{lagr}).
When  there is no  $A$-coupling  term  ($q=0$)
the electric and
magnetic charges are orthogonal, i.e. they are associated with
gauge fields originating from   two different compactified directions.
The $A$-coupling induces new electric charges,  but only
 in a  left-moving direction:
 it leads to  non-zero left-moving electric charge $Q_{L\,1}\equiv
{1\over 2} (Q^{(1)}_1-Q^{(2)}_1)=q$
 along the magnetic charge direction.  The
 right-moving charges, i.e.
 $P_{R\,n}\equiv {1\over 2} (P^{(1)}_n+P^{(2)}_n)$ and $Q_{Rn}\equiv{1\over 2}
(Q^{(1)}_n+Q^{(2)}_n)$,  still remain orthogonal.

The explicit form of the space-time metric function
$\lambda$, the
dilaton $\Phi'$, the axion $\Psi$ and the moduli $G_{mn}$, $B_{mn}$
of the
internal  two-torus is: \begin{equation}
\lambda={{r^2}\over{\bigg[(r+Q_1)(r+Q_2)(r+P_1)(r+P_2)
 -  q^2 [r + {1\over 2}(P_1 +P_2)]^2\bigg]^{1\over 2}}}\   ,
\label{meto} \end{equation}
 \begin{equation}
e^{2\Phi'}={{(r+P_1)(r+P_2)}\over{\bigg[(r+Q_1)(r+Q_2)(r+P_1)(r+P_2)
 -    q^2 [r + {1\over 2}(P_1 +P_2)]^2\bigg]^{1\over 2}}} \ ,
\label{dilo}\end{equation}
\begin{equation}
\Psi={q (P_2 - P_1) \over  2 (r+P_1) (r + P_2) } \   ,
\label{axo}
\end{equation}
\begin{equation}   G_{11}={{r+P_2}\over{r+P_1}},  \ \ 
G_{22}={{r+Q_1}\over{r+Q_2}}, \ \
 G_{12}=-B_{12}= {{q [r+{1\over 2} (P_1+P_2)]}\over{(r+Q_2)(r+P_1)}}.
   \label{modo}
\end{equation}
Note that this one-center solution is written  with the  following choice of
the asymptotic  ($r\to \infty$)  values for  the fields:
$\Phi'_\infty=\Psi_\infty=G_{12\,\infty}=B_{12\,\infty}=0$ and
$G_{11\,\infty}=G_{22\,\infty}=1$. Solutions with other aymptotic
values of the  axion-dilaton and moduli fields are related to
 this one by a  particular $SL(2,R)$
($S$-duality) and $O(2,2)$ ($T$-duality of two-torus)
transformations, respectively.

Regular solutions, i.e. solutions with event horizons,
are determined by
choosing  the four  parameters  $P_{1,2},Q_{1,2}$ to be
{\it positive}:
\begin{equation}
P_1>0, \ \ \ P_2>0, \ \ \ Q_1>0,\ \ \ Q_2>0,  \
\label{chconsp}
\end{equation}
 and $q$  satisfying the following constraint:
\begin{equation}
Q_1Q_2-q^2>0,\ \ \ \  \ \ (Q_1Q_2-q^2)P_1P_2- {1\over 4} q^2(P_1-P_2)^2>0.
\label{chcons}
\end{equation}
Regular solutions,\footnote{The space-time properties of regular
solutions with $q=0$ were studied in Ref. \cite{CY}.}
 i.e. those satisfying   all the  inequalities (\ref{chconsp}),(\ref{chcons}),
 have an event horizon at $r=0$ and a
time-like  singularity  at  negative $r=r_{sing}$ ($-{\rm min}\{P_1,P_2,Q_1,Q_2\}
<r_{sing}<0$
 for $q\not=0$   and  $r_{sing}=-{\rm min}\{P_1,P_2,Q_1,Q_2\}$ for $q=0$), i.e.
 the global  space-time is that  of
extreme Reissner-Nordstr\"om black holes.
In the case when any  of the charge
combinations in  (\ref{chconsp}),(\ref{chcons})  is  zero,
 the singularity is null and located  at $r=0$, i.e.  the
horizon and the singularity  coincide.  In  the case of only one non-zero
parameter in (\ref{chconsp}) the singularity at $r=0$ is  naked.
When at least one of  the charge combinations in (\ref{chconsp}),(\ref{chcons})
becomes negative the solutions are singular with a  naked singularity
at $r>0$.

However, we would like  to emphasize
that for small $r$  the effective four-dimensional
description breaks down and the  solution  becomes effectively
six-dimensional. Therefore the  nature of
 singularities should be re-addressed from the  point of view of the
 six-dimensional  string  theory,  using string-frame metric
 (cf. \cite{GHT}).
For $r\geq 0$,  the  regular   solutions
 satisfying (\ref{chconsp}),(\ref{chcons})  are
 always  {\it nonsingular}.
This is a  reflection of the regularity of the underlying
six-dimensional conformal  $\sigma$-model  discussed in  Section II.

In the following we shall concentrate on  regular solutions.
The asymptotic value of the metric coefficient $\lambda$  (\ref{meto})
is of the form:
\begin{equation}
\lambda=1-{M_{ADM}\over {2 r}} +{\cal O}({r^{-2}}),
\label{masse}
\end{equation}
where the  ADM mass
\begin{equation}
M_{ADM} = Q_1 +Q_2 +P_1 +P_2 \
\label{mass}
\end{equation}
 does not depend on $q$.
It saturates the Bogomol'nyi bound
\cite{CY,DLR}  and  corresponds to the BPS-saturated state that preserves
${1\over 4}$ of the original $N=4$ target space supersymmetry.

Scalar fields (\ref{dilo})-(\ref{modo}) have the following asymptotic behaviour
\begin{equation}
e^{2\Phi'}=1+{{ (P_1+P_2-Q_1-Q_2)}\over {2r}} + {\cal O}(r^{-2}),
\ \ \ \ \   \Psi ={{q(P_2- P_1)}\over {2r^2}} +{\cal O} ( r^{-3}),
\label{axia}
\end{equation}
\begin{equation}
G_{11}=1 +{{(P_2-P_1)}\over{r}} + {\cal O}(r^{-2}),
   \ \ \ \ G_{22}=1+ {{(Q_1-Q_2)}\over{r}}  + {\cal O}(r^{-2}),
\label{dmoda}
\end{equation}
\begin{equation}
 G_{12}=-B_{12}= {{q}\over{r}} + {\cal O}(r^{-2}).
   \label{moda}
\end{equation}
Note that the dilaton and all the two-torus moduli have non-zero scalar
charges, while the axion  charge is zero.

The area of the event  horizon, i.e. ${\bf A}\equiv 4\pi(\lambda^{-1}
r^{2})_{r=0}$, is
 \begin{equation}
{\bf A}
= 4\pi \bigg[ (Q_1Q_2-q^2)P_1P_2 - {1\over 4}  q^2(P_1
-P_2)^2\bigg]^{1\over 2}
.  \label{mes}
\end{equation}
For $q\not= 0$ the  area  is decreased   compared  to its value for  $q=0$.

At the horizon  at $r=0$  all the scalars, i.e. 
 the axion $\Psi$ (\ref{axo}), the dilaton $\Phi'$ (\ref{dilo}) and
 the moduli (\ref{modo}) are constant.
Note that unlike
for  pure electric or pure magnetic configurations, where
 $\Phi'$  grows  either at small or at large distances,  here $e^{\Phi'}$
can  be
chosen   to be small in the whole region  $r\geq 0$.
%, provided  $(P_1P_2)^2 < Q_1Q_2P_1P_2- {1\over 4}q^2(P_1+P_2)^2$.

The $T$-self-dual  case with
$Q_1=Q_2=Q, \ P_1=P_2=P$  deserves  a special discussion. 
In this case the   moduli $G_{11}$
and $G_{22}$  (\ref{modo}) remain constant.
For $q=0$  and  $Q=P$  it
corresponds to the  extreme Reissner-Nordstr\"om-type  dyonic black hole
with all scalar fields  being constant
\cite{CY,US}.
In this case  the six-dimensional  $\sigma$-model  (\ref{lagr}),(\ref{kkk})
 takes  a particularly   simple
form  discussed in \cite{US}.
 For  $Q=P$ the  six-dimensional dilaton (\ref{diit})
is   constant, but if $q\not=0$  the moduli  $G_{12}=-B_{12}$,
 and thus also  the  four-dimensional dilaton $\Phi'$, are
no longer constant.
 The
area  of the horizon  in this case  is given by
${\bf A}  = 4\pi P  \sqrt{ Q^2 -  q^2 }$.

%%%%%%%%%%%%%%%%%%%%%%%%%%%%%%%%%%%%%%%%%%%%%%%%%%%%%%%%%%

%%%%%%%%%%%%%%%%%%%%%%%%%%%%%%%%%%%%%
\section{All BPS saturated  static black hole solutions of  heterotic string
compactified on  six-torus}
 %%%%%%%%%%%%%%%%%%%%%%%%%%%%%%%%%%%%%%%%%%%

The  five-parameter solution obtained in Section III
turns out  to be a generating
solution for {\it  all}   static spherically-symmetric BPS-saturated
configurations of the four-dimensional heterotic string compactified on
six-torus.
 These solutions can be obtained by applying  a subset of
 $T$-duality ($O(6,22)$)  and $S$-duality
($SL(2,R)$)) transformations to the generating solution.
 It should be noted, however, that
 while the generating solution  is described by
an exact conformal $\sigma$-model  (\ref{lagr}), (\ref{latra}),  
these more general BPS-saturated  backgrounds
are  guaranteed  only to be
solutions of  the leading-order effective
 string equations.
Indeed,  in contrast to  the $T$-duality transformations,
the $S$-duality
transformations  do not,  in general,  map  one conformal $\sigma$-model into 
another.
%%%%%%%%%%%%%%%%%%%%%%%%%%%%%%%%%%%%%%%%%%%%%%%
\subsection{Effective four-dimensional action}
%%%%%%%%%%%%%%%%%%%%%%%%%%%%%%%%%%%%%%%%%%%%%%%%%%%%%

The effective four-dimensional  heterotic string compactified on
six-torus has $N=4$  supersymmetry. The bosonic part
of the leading term in the
 effective action  has the following form (for a review see
 \cite{SEN1} and references therein):
$$
S={1\over 16 \pi G_N} \int {\rm d}^4 x
\sqrt{-g}
\bigg[{ R(g)} -2\partial_{\mu} \Phi' \partial^{\mu} \Phi'
- {{1}\over {12}}{\rm e}^{-4\Phi'} {H}^{'}_{\mu\nu\rho}{H}^{'\mu\nu\rho}
$$
\begin{equation}
  - \  {{1}\over 4}{\rm e}^{-2\Phi'}{\cal F}^i_{\mu\nu}(LML)_{ij}
{\cal F}^{j\mu\nu} +
 {1\over 8}{\rm Tr}(\partial_{\mu} ML \partial^{\mu}ML)\bigg]\ .
\label{string}
\end{equation}
 The action (\ref{string}) depends on
 massless four-dimensional bosonic fields, which are determined in
 terms of the following dimensionally reduced  ten-dimensional fields:
Zehnbein
   $\hat{E}^A_M$,  dilaton $\Phi$,
two-form field ${B}_{MN}$ and $U(1)^{16}$ gauge fields ${A}^I_M$  ($M,N=0,...,
9$; 
$I=1,...,16$). The Ansatz for the Zehnbein  is  of the form
$$\hat{E}^A_M=\left ( \matrix{e^{\Phi'}e^{\alpha}_{\mu} &
A^{(1)\,m}_{\mu}e^a_m \cr 0 & e^a_m} \right )\ , $$
  where $A^{(1)m}_\mu$
($m=1,...,6;  \  \mu=0,...,3$) are Kaluza-Klein $U(1)$
 gauge fields, $\Phi'=\Phi- {1\over 2} {\rm
ln}{\rm  det}{e^a_m}$  is the four-dimensional dilaton field, and
 $g_{\mu\nu}=e^{\alpha}_{\mu}e^\alpha _\nu$  is the
 Einstein frame  metric.
 Other components of   28 $U(1)$ gauge fields
${\cal A}^i_{\mu} \equiv (A^{(1)\, m}_{\mu}, A^{(2)}_{\mu\, m},
A^{(3)\, I}_{\mu})$  are defined as $A^{(2)}_{\mu\,m} \equiv {B}_{\mu\,m}
+ {B}_{mn}A^{(1)\,n}_{\mu} + {1\over 2}a^I_m A^{(3)\,I}_{\mu}$,
$A^{(3)\,I}_{\mu} \equiv {A}^I_{\mu} - a^I_m A^{(1)\,m}_{\mu}$ with the
field strengths ${\cal F}^i_{\mu\nu} = \partial_{\mu} {\cal A}^i_{\nu}
-  \partial_{\nu} {\cal A}^i_{\mu}$. The two-form field  with the field
strength
$H^{'}_{\mu\nu\rho} =3(\partial_{[\mu} B_{\nu\rho]} + {1\over 2}{\cal A}_{[\mu} L
{\cal F}_{\nu\rho]})$
 is  equivalent to a
pseudo-scalar (the axion)  $\Psi$ through the duality transformation
$H^{'\mu\nu\rho} = {{e^{4\Phi'}}\over {\sqrt{-g} }}
\varepsilon^{\mu\nu\rho\sigma}\partial_{\sigma}\Psi$.
  A symmetric $O(6,22)$
matrix $M$ of scalar (moduli) fields  can be expressed in terms of
the following $O(6,22)$ matrix $V $   \cite{MS}
\begin{equation}
M = V^T V, \ \ \  \ \ \ V  =
\left ( \matrix{E^{-1} & E^{-1}C & E^{-1}a^T \cr
0 & E & 0 \cr 0 & a & I_{16}} \right )  ,
\label{mviel}
\end{equation}
 where $E \equiv [e^a_m]$ (the Sechsbein of the internal
metric $G_{mn}$), $C \equiv [{1\over 2}
{A}^I_m {A}^I_n + {B}_{mn}]$ and $a \equiv [{A}^I_m]$.
$V$ plays a role of a Vielbein in the $O(6,22)$ target space.

The four-dimensional  effective action is invariant under the $O(6,22)$
transformations ($T$-duality) \cite{MS,SEN1}:
\begin{equation}
M \to \Omega M \Omega^T ,\ \ \ {\cal A}^i_{\mu} \to \Omega_{ij}
{\cal A}^j_{\mu}, \ \ \ g_{\mu\nu} \to g_{\mu\nu}, \ \ \ S \to S ,
\label{tdual}
\end{equation}
where $S \equiv \Psi + i e^{-2\Phi'} $  and  $\Omega \in O(6,22)$
is an $O(6,22)$ invariant matrix,
\begin{equation}
\Omega^T L \Omega = L \ , \ \ \ \ \
L =
\left ( \matrix{0 & I_6& 0\cr
I_6 & 0& 0 \cr 0 & 0 & - I_{16}} \right ).
\label{mvie}
\end{equation}
   In addition, the
corresponding equations of motion and Bianchi identities are
invariant under the $SL(2,R)$ transformations ($S$-duality) \cite{SEN1}:
\begin{equation}
S \to {{aS+b}\over{cS+d}},\ \ M\to M ,\ \ g_{\mu\nu}\to g_{\mu\nu},\ \
{\cal F}^i_{\mu\nu} \to
(c\Psi + d){\cal F}^i_{\mu\nu} + ce^{-2\Phi'} (ML)_{ij}
\tilde{\cal F}^j_{\mu\nu},
\label{sdual}
\end{equation}
where  $\tilde{\cal F}^{i\,\mu\nu} =
{1\over 2\sqrt{-g}} \varepsilon^{\mu\nu\rho\sigma}
{\cal F}^i_{\rho\sigma}$ and $a,b,c,d \in R$ satisfy $ad-bc=1$.
At the quantum level,  the parameters
of both $T$- and $S$-duality transformations  become
integer-valued.

%%%%%%%%%%%%%%%%%%%%%%%%%%%%%%%%%%%%%%%%%%%%%%%
\subsection{General class of  dyonic  solutions generated by  duality
transformations}
%%%%%%%%%%%%%%%%%%%%%%%%%%%%%%%%%%%%%%%%%%%%%%%
Static spherically-symmetric solutions corresponding to
 the effective action
(\ref{string}) are described
 by the   Ansatz  (\ref{eee})
for the
four-dimensional
space-time metric, the  dilaton-axion  field $S$  and the  moduli fields $M$,
which depend only
on the radial coordinate $r$,
 and by 28
   electric and 28 magnetic    $U(1)$ gauge fields.
 The  Maxwell's equations  and the  Bianchi
identities  determine the components of the $U(1)$ field strength to be
\begin{equation}
\sqrt{2}{\cal F}^i_{tr} = 4 G_N {e^{2\Phi'(r)} \lambda(r)\over r^2}
M_{ij}(\alpha_j +  \Psi \beta_j)
,  \ \ \ \ \  \ \ 
\sqrt{2}{\cal F}^i_{\theta\phi} = L_{ij}\beta_j\, {\rm sin}\,\theta,
 \label{elecmag}
\end{equation}
 which are expressed  in terms of the conserved charge vectors $\vec \alpha$
and
$\vec \beta$.
The  electric and magnetic  charges
\begin{equation}
\vec P^T\equiv (P^{(1)}_m; P^{(2)}_m; P^{(3)}_I), \ \ \ \ \ \
\vec Q^T\equiv(Q^{(1)}_m; Q^{(2)}_m; Q^{(3)}_I) ,
\label{ccc}
\end{equation}
are  related
 to the charge vectors $\vec \alpha$ and $\vec \beta$  in the following
  way \cite{WITTENIII}:
\begin{equation}
\sqrt{2}Q_i = e^{2\Phi'_{\infty}}
M_{ij\,\infty}(\alpha_j + \Psi_{\infty}\beta_j),
\ \ \ \ \ \ \ 
\sqrt{2}P_i=L_{ij}{\beta}_j,  \label{charges}
\end{equation}
where  the subscript
 $\infty$ refers to the asymptotic ($r\to\infty$)
  values of $M$ and $\Psi$  
and we have assumed that $\alpha'=2, \ G_N= {1 \over 8} \alpha' e^{2\Phi'_\infty} =
{1 \over 4} e^{2\Phi'_\infty}$
%,   \ ${\Phi'(\infty)} =0$
.

The
equations of motion
are invariant
under both $T$- and $S$-duality transformations. Therefore,
 one can generate new
supersymmetric solutions by applying $O(6,22)$ and $SL(2,R)$
transformations to  some  known  solution.
This is the technique
\cite{sen,CY,CYIII}\footnote{Analogous techniques, employing the symmetries
 of the effective four-dimensional  as well as
three-dimensional action of the ($4+n$)-dimensional Abelian Kaluza-Klein
theory were used \cite{CYIV} to obtain all static spherically-symmetric  black
holes in that theory.} which was used previously
to obtain a general class of  BPS-saturated  backgrounds.
In particular, starting with
 the four-parameter  BPS-saturated
solution one  finds  \cite{CY}  a  general class of  BPS-saturated
black hole solutions
with 28 electric and 28 magnetic charges subject  to one constraint.
 Here we follow the same procedure to obtain the
most general BPS-saturated solution
starting with  the five-parameter solution of
Section III.
We shall consider  regular solutions with event horizons.
 In particular, we shall determine the expression for
  the ADM mass formula
and  for the area of the event horizon
for the most general BPS-saturated
 configuration in this class.

Without loss of generality one can bring \cite{SEN1}
arbitrary asymptotic values of the moduli and axion-dilaton fields to the form
 $M_{\infty}=I$
and $S_{\infty}=i$
by performing the following $O(6,22)$
and $SL(2,R)$
 transformations:
\begin{equation}
M_{\infty} \to \hat{\Omega}M_{\infty}\hat{\Omega}^T=I ,\ \  \ \
  S_{\infty} \to (aS_{\infty}+b)/d = i.
\label{asympt}
\end{equation}
Here
$\hat{\Omega} \in O(6,22)$, $ad=1$, and in  quantized theory
the charge lattice vectors will belong to
 the new transformed lattice.
Then the subsets of $O(6,22)$ and $SL(2,R)$ transformations
that preserve the above new asymptotic values of $M_\infty$  and $S_\infty$
are
$O(6)\times O(22)$ and $SO(2)$  transformations, respectively.
Note  that configurations obtained in that manner
have the same four-dimensional space-time structure and thus the same
singularity and thermal properties as the generating solution.
To find  solutions
with arbitrary asymptotic values of $M$ and $S$ one has to undo
the above transformations.

The four-dimensional black hole  background corresponding to   the
solitonic string solution  described in
  Sections II and  III  is  parameterised  by the
two magnetic   $P^{(1,2)}_1$ and  the four
electric
$Q^{(1,2)}_2$ and $Q^{(1)}_1=-Q^{(2)}_1\equiv q$ \  charges, i.e.
$$
{\vec Q}^T = (q,Q_1,0, ..., 0;-q, Q_2,0, ..., 0;0, ..., 0), \  \ \
{\vec P}^T= (P_1,0, ..., 0; P_2,0,  ..., 0;0, ..., 0).
$$
In Section III the
asymptotic values of the
moduli and the dilaton-axion fields  were already  chosen to be of the form
$M_\infty =I$, $S_\infty=i$. This background
 can  now  be used as a   generating
 solution for the most general  set of  solutions in this class.

As a first step,  one  applies
 a subset of $O(6)\times O(22) \subset
O(6,22)$
 transformations  which correspond to $SO(6)/SO(4)$
transformations with 9
parameters and $SO(22)/SO(20)$ transformations with 41 parameters,
which,  along with  $5$
  original charges,  give
 a configuration with $56$ (28 electric ${\vec Q}$ and
28 magnetic ${\vec P}$) charges subject to one constraint.
 After  one has undone the transformation (\ref{asympt}), so that $M_\infty$
and
$S_\infty$ become arbitrary,  this
constraint can be cast into the following $O(6,22)$ ($T$-duality) invariant  
form:
\begin{equation}
{\vec P}^T{\cal M}_{+}{\vec Q}=0\ , \ \ \ \
 \ {\cal M}_{\pm} \equiv LM_{\infty}L\pm L\ .
\label{gencon}
\end{equation}
The ADM mass  (\ref{mass}) of  the generating solution
 can be written in the following
$O(6,22)$  invariant form:
\begin{equation}
M^2_{ADM} ={\rm e}^{-4\Phi_\infty'}\bigg[({\vec Q}^T{\cal M}_{+}{\vec
Q})^{1\over 2} + 
({\vec P}^T{\cal M}_{+}{\vec P})^{1\over 2}\bigg]^2 .
\label{TBPS}
\end{equation}
The area of the event horizon $\bf A$   (\ref{mes}) for the
regular  generating solution
can also be cast  in  the  $O(6,22)$ invariant form:
\begin{equation}
{\bf A}=2\pi \bigg[({\vec P}^TL{\vec P})({\vec
Q}^TL{\vec
Q})- {1\over 4} ({\vec P}^T{\cal M}_-{\vec Q})^2\bigg]^{1\over 2}.
\label{tent}\end{equation}
Using the  charge constraint (\ref{gencon}) one can replace  the term
$({\vec P}^T{\cal M}_{-}{\vec Q})$  in (\ref{tent}) by $2({\vec P}^T(LM_{\infty}L){\vec
Q})$ and represent  (\ref{tent}) in the following
% $O(6,22)$ ($T$-duality) invariant
$SL(2,R)$ ($S$-duality) invariant form:
\begin{equation}
{\bf A}=
2\pi \bigg[({\vec P}^TL{\vec P})({\vec Q}^TL{\vec Q})-
 ({\vec P}^T{(LM_{\infty}L)}{\vec Q})^2\bigg]^{1\over 2}.
\label{Tent}
\end{equation}
The  subsequent  $SO(2) \subset SL(2,R)$ transformation provides  one
more parameter, 
which removes the
charge constraint (\ref{gencon}). The most  general configuration
in this class   has  then 56
parameters
specified by {\it unconstrained} 28 electric ${\vec
Q}$ and 28 magnetic ${\vec P}$ charges.
 This configuration thus corresponds to
the most general  spherically-symmetric
static BPS-saturated black hole
solution consistent with the  no-hair theorem.

The $SO(2) \subset SL(2,R)$  transformation  allows one
to write the ADM mass formula
in the following $O(6,22)$ and $SL(2,R)$ invariant
 form \cite{CY,DLR}:
\begin{equation}
M^2_{ADM} = e^{-4\Phi'_\infty} \bigg({\vec Q}^T {\cal M}_+ {\vec Q}
+{\vec P}^T {\cal M}_+ {\vec P} 
+ \ 2\left [({\vec P}^T{\cal M}_+{\vec P})
({\vec Q}^T{\cal M}_+{\vec Q})-({\vec P}^T{\cal M}_+
{\vec Q})^2\right]^{1\over 2}\bigg).
\label{Bogmass}
\end{equation}
Note that when the magnetic  and electric  charges
are parallel in the $SO(6,22)$ sense, i.e. $\vec{P}\propto \vec{Q}$,
the ADM mass (\ref{Bogmass}) corresponds to the mass of the
BPS-saturated black holes  which  preserve ${1\over 2}$ of $N=4$ supersymmetry
(see, e.g., \cite{SEN1}).
        In the case when the magnetic and
electric charges are not parallel, the mass is larger and the
configurations preserve $1\over 4$ of $N=4$ supersymmetry.

The  area of the event horizon (\ref{Tent}) is already invariant under the
$SL(2,R)$ transformations and  thus remains of the same form.
 The general expression for the  area
 reduces to the special form when the charge configurations are constrained.
Regular configurations  with  ${\vec P}\propto {\vec Q}$, i.e.
BPS-saturated states
which  preserve ${1\over 2}$ of $N=4$ supersymmetry, have zero  area of the
event horizon.

 Another example is provided by  the most general  solutions  with {\it
zero} axion
 \cite{CY}.
 Those are  backgrounds   obtained from the
  four-parameter generating  solution  with $q=0$ by applying
 a subset
 of  $O(6,22)$ transformations. They are  specified  by  28
 electric  and  28  magnetic    charges  subject to {\it two} constraints:
 ${\vec P}^T{\cal M}_{+}{\vec Q}=0$
{ and} ${\vec P}^T{\cal M}_{-}{\vec Q}=0$ \cite{CY}. Thus,
in this case ${\vec P}^T(LM_\infty L){\vec Q}=0$, and only
the first term  in the expression (\ref{Tent}) is present,
 as pointed out in \cite{LW}.
In general,  the area  of the event horizon is
 {\it decreased} by  an  additional positive definite
term which  measures the  orthogonality
of the magnetic and electric charge vectors.

%%%%%%%%%%%%%%%%%%%%%%%%%%%%%%%%%%%%%%%%%%%%%%%%%%%%
\subsection{ADM mass and area of  horizon in terms of conserved charges}
%%%%%%%%%%%%%%%%%%%%%%%%%%%%%%%%%%%

As the last step,  we  can  express   the ADM mass formula (\ref{Bogmass}) and
the area
of the event horizon (\ref{Tent}) in terms of the
conserved   electric $\vec\alpha$ and magnetic
 $\vec\beta$ charge vectors, thus   allowing  to study
 their dependence
on the asymptotic
 values of the the axion-dilaton  $S_\infty$ and  moduli $M_\infty$ fields.
Since $\vec P$ and $\vec Q$ are related to the conserved charge vectors
$\vec\alpha$  and $\vec\beta$ through (\ref{charges}),  the ADM  mass formula
(\ref{Bogmass}) can  be written as:
\begin{equation}
M_{ADM}^2 = 
{1\over 2 }{\vec {\tilde\alpha}^T} \mu_+ {\vec {\tilde\alpha}}
+ {{1\over 2} }e^{-4\Phi'_\infty} {\vec \beta}^T \mu_+{\vec \beta} + e^{-2\Phi'_\infty} \left [({\vec \beta}^T\mu_+{\vec \beta})
({\vec \alpha}^T\mu_+{\vec \alpha})-({\vec \beta}^T\mu_+
{\vec \alpha})^2\right]^{1\over 2},
\label{Bogmasspp}
\end{equation}
where ${\vec {\tilde \alpha}}\equiv{\vec \alpha}+\Psi_{\infty}{\vec\beta}$ and
$\mu_{\pm}\equiv M_{\infty}\pm L$. 
Similarly,  the area of the event horizon (\ref{Tent}) 
and  thus the black hole entropy 
can be  represented
as:\footnote{The four-dimensional dilaton value at the event horizon
is also a moduli- and coupling-independent quantity: $e^{2\Phi'(0)}={\vec
\beta}^TL{\vec\beta}$ $[({\vec \beta}^T L {\vec \beta})
({\vec {\alpha}}^T L{\vec {\alpha}})-({\vec \beta}^T L  {\vec
{\alpha}})^2]^{-{1\over 2}}$.}
\begin{equation}
{\bf A}=\pi e^{2\Phi'_\infty}\bigg[({\vec \beta}^T L {\vec \beta})
({\vec {\alpha}}^T L{\vec {\alpha}})-({\vec \beta}^T L  {\vec
{\alpha}})^2\bigg]^{1\over 2} \ , 
%\label{genetpp}
%\end{equation}
%\begin{equation}
 \ \  {\bf S} = {{\bf A} \over 4G_N}
% = 
%\pi \bigg[({\vec \beta}^T L {\vec \beta})
%({\vec {\alpha}}^T L{\vec {\alpha}})-({\vec \beta}^T L  {\vec
%{\alpha}})^2\bigg]^{1\over 2} \ . 
\label{genentpp}
\end{equation}

Both the ADM mass (\ref{Bogmasspp}) 
and  the entropy (\ref{genentpp}) can be cast in the {\it manifestly} 
$T$- {\it and} $S$-duality  invariant form as:\footnote{We would like to thank C. Hull for discussions on that point.}
\begin{equation}
M_{ADM}^2 = (8G_N)^{-1}\left(
{} {\cal M}_{\infty\, ab}({\vec v}^{a\, T}\mu_+ 
{\vec v}^b) +\left[{ 2}{\cal L}_{ac}{\cal L}_{bd}({\vec v}^{a\, 
T}\mu_+{\vec 
v}^b) ({\vec v}^{c\, T}\mu_+{\vec v}^d)\right]^{1\over 2}\right),
\label{Bogmassp}
\end{equation}
%\begin{equation}
%{\bf A}=4\pi G_N \bigg[{\cal L}_{ac}{\cal L}_{bd}({\vec v}^{a\ T} L {\vec %v}^{b})
%({\vec v}^{c\ T} L{\vec v}^{d})\bigg]^{1\over 2} \ , 
%\label{genetp}
%\end{equation}
\begin{equation}
 {\bf S} = {{\bf A} \over 4G_N}=\pi  \bigg[{1\over 2}{\cal L}_{ac}{\cal L}_{bd}({\vec v}^{a\, T} L {\vec v}^{b})
({\vec v}^{c\, T} L{\vec v}^{d})\bigg]^{1\over 2}\ .
\label{genentp}
\end{equation}
Here  ${\cal M}_{\infty}$  is the asympotic value of the axion-dilaton  
matrix
${\cal M}$, and ${\cal L}$ is the $SL(2,R)$ invariant matrix. ${\cal M}$ and ${\cal L}$  are of the following form:
\begin{equation}
{\cal M}={\rm e}^{2\Phi'}
\left ( \matrix{1 &  \Psi\cr
\Psi& \Psi^2+{\rm e}^{-4\Phi '}} \right ) \ \ ,\ \
{\cal L} =
\left ( \matrix{0 & 1\cr
-1& 0} \right ),
\label{mlsl}
\end{equation}
and the (quantised) charge vector ${\vec v}^T= ({\vec v}^{1\, T}, {\vec 
v}^{2\, T})\equiv ({\vec \alpha}^T, {\vec \beta}^T)$.
${\cal M}$  and ${\vec v}$ transform under the $SL(2,R)$ transformation 
$\omega =\left(\matrix{d&c\cr b&a}\right)$  $(ad-bc=1)$ as\cite{SEN1}
\begin{equation}
{\cal M }\to \omega {\cal M }\omega^T ,\ \ \ {\vec v} \to  {\cal L}\omega{\cal L}^T{\vec v}.
\label{trans}
\end{equation}
An important observation is 
that while for fixed 
values of the charge vectors  $\vec\alpha$ and $\vec \beta$  the ADM mass
(\ref{Bogmassp}) changes  under   the
variations of the  moduli and   string coupling,
 {\it the  entropy 
 remains the same
 as one moves in the  moduli and  coupling  space}.
The  fact
the  entropy
is an invariant quantity   is
consistent with  the  expectation  that the internal structure of  a
BPS-saturated black hole
 should not change under
variations
 of
 moduli and couplings.
 
 This invariance
indicates   \cite{LW}  that  the classical entropy  may   have
a statistical interpretation
in terms of a  number
of degenerate black-hole configurations:  being an  integer such a  number
 would  not change  under adiabatic variations
 of moduli and axion-dilaton couplings.
Indeed,  the combination
of charges $({\vec \beta}^T L {\vec \beta})
({\vec {\alpha}}^T L{\vec {\alpha}})-({\vec \beta}^T L  {\vec
{\alpha}})^2$ which appears in
(\ref{genentp})
is expected to be an  (even) integer.
Following \cite{SEN1},
one may attempt to justify this  by using
the
analogy with the level matching condition for the elementary
 BPS-saturated string states of   toroidally compactified heterotic string and
the  Dirac-Schwinger-Zwanziger-Witten (DSZW) \cite{WITTENIII}
 quantisation condition.
In the  case of the generating solution   described
by the conformal model  discussed in Section II,
the  quantisation of charges
is implied  by the consideration
of the conformal theory corresponding  to the throat region (cf. Section II.B).

Note that the  purely electric BPS-saturated black holes
 preserve $1\over 2$ of $N=4$ supersymmetry
and  have the same quantum numbers \cite{CMP,dab}
as the  elementary   BPS-saturated string states with
no excitations in the right sector
 ($N_R={1\over 2}$). In the electric  case the
quantised    charge vector $\vec\alpha$  is constrained to lie on an even
self-dual
   lattice  with
   the     norm
\cite{sen}
\begin{equation}
{\vec \alpha}^TL{\vec\alpha}=2N_L-2=-2,0, 2, ... \ .
\label{chl}
\end{equation}
 The  DSZW  charge quantisation condition then  implies
an analogous  constraint for
${\vec \beta}^TL{\vec\beta}$.
 
 The necessary conditions  for a  BPS-saturated  configuration to be regular
 with  a   non-zero area of the event horizon   are
\begin{equation}
{\vec \alpha }^TL{\vec\alpha }> 0  \ ,\  \ \ \
{\vec \beta}^TL{\vec\beta}> 0\ , \  \ \  \
({\vec \beta}^T L {\vec \beta})
({\vec {\alpha}}^T L{\vec {\alpha}})-({\vec \beta}^T L  {\vec
{\alpha}})^2 > 0.
\label{norm}
\end{equation}
The latter constraint  becomes an equality
  for the  BPS-saturated configurations  preserving
$1\over 2$ of $N=4$
supersymmetry
(i.e., for quantized charges,   when
${\vec \alpha}\propto {\vec \beta}$ with magnetic and
electric charge vector components being co-prime integers
\cite{SEN1}).
%%%%%%%%%%%%%%%%%%%%%%%%%%%%%%%%%%%%%%%%%%%%%%%%%%%%%%%%

%%%%%%%%%%%%%%%%%%%%%%%%%%%%%%%%%%%%%%%%
%%%%%%%%%%%%%%%%%%%%%%%%%%%%%%%%%%%%%
\section{String
 origin of dyonic black hole entropy}
%%%%%%%%%%%%%%%%%%%%%%%%%%%%%%%%%%%%%%%%%
%%%%%%%%%%%%%%%%%%%%%%%%%%%%%%%%%%%%%%%%%%%%%%%%
One of the motivations behind the above  discussion   of
 the five-parameter dyonic static spherically-symmetric BPS-saturated  black
 hole  as a  four-dimensional
`image' of an exact  solitonic string solution in six dimensions
is to try to use information about the underlying
 conformal  theory
in order to give a statistical interpretation to the black hole
entropy. The  aim is to  extend  and
amplify  the  recent    interesting  proposal \cite{LW}
along these lines  which generalise  an earlier suggestion \cite{sen}
(see also \cite{suss}). The discussion  in \cite{LW} was based on
a subset of BPS-saturated  dyonic black holes of heterotic string on six-torus,
 namely  the most  general configurations  with zero
axion. Those can be obtained by $T$-duality transformations on the
four-parameter,  i.e. $q=0$,  generating solution
 considered  in \cite{CY,US}.
The  expressions for the area  and 
the  entropy  of  these solutions correspond  to  a special case of
(\ref{genentp})  with  ${\vec \beta}^TL{\vec \alpha}=0$. We expect
 that the inclusion
of the new parameter $q$ should
reveal certain more  general
aspects  of the relation between the entropy and  the degeneracy
of  black hole  states.

Just as in the case of   purely  electric BPS-saturated black holes
whose entropy  may be  qualitatively understood as  being a consequence 
of  degeneracy of states 
originating from  
oscillations of the fundamental string \cite{CMP,dab},
one would like  to  explain  the entropy  of the dyonic
BPS-saturated black holes 
in terms of degeneracy of states 
originating from 
small-scale oscillations of  the underlying
six-dimensional string soliton.

It was  already emphasized in Section I that
an important advantage of the  dyonic black holes, studied in Sections III
and IV,   over the electric ones, considered in \cite{sen,peet,CMP,dab},
is that the magnetic charges provide a short-distance `regularisation'
 of the metric and that
 the dilaton is approximately constant.\footnote{The presence of
 the magnetic charges provides  a
`regularisation'   making it unnecessary
to resort to  `stretched horizon' considerations   used in  the
case of purely electric extreme black holes \cite{sen,peet}.
There is  a certain analogy   between the present dyonic model
and the conjectured  modification (by world-sheet $\alpha'$-corrections)
of the purely electric model   \cite{sen}.
 Consideration of the 
stretched horizon  region  corresponds effectively to a shift $r\to r+ \sqrt{\alpha'}$
in (part of) the metric. Analogous regularisation of singularities
of the fundamental string and extreme electric black hole solutions
was suggested in \cite{ME}.
 Turning on magnetic charges  $P_1 =P_2=P$ can
 be represented as a  replacement of   one  of the $r^2$-factors
in the metric  by
$(r + P)^2$.
 Then the $r=0$ region becomes
non-singular provided all four charges $(Q_n,P_n)$ are non-vanishing.}
As a result, one may hope to understand the
statistical origin of the black hole entropy
starting directly  from string theory  and
 using only  semiclassical  considerations.

 %%%%%%%%%%%%%%%%%%%%%%%%%%%%%%%%%%%%%%%%%%
\subsection{Statistical entropy and magnetic renormalisation of $\alpha'$}
%%%%%%%%%%%%%%%%%%%%%%%%%%%%%%%%%%%%%%%%%

Let us consider the case when all the charges are large and the magnetic
charges $P_1,P_2$ are
 approximately equal. Since the value of the dilaton at  the  $r=0$
horizon
is $e^{2\Phi_0}= e^{2\Phi_\infty }  P_2Q_2^{-1} $  (cf. (\ref{diit}))
to get a small value for  the string coupling in the horizon region
one needs  to assume  that $Q_2$ is also large.
Then the expression for the thermodynamic black-hole
entropy,
proportional to the area of the horizon
(\ref{mes}), is of the form:\footnote{The assumption
$P_1\approx P_2$ implies that the
second term under the square root
in (\ref{mes}),(\ref{genentp}) can be neglected.}
\begin{equation}
{\bf S} = {{\bf A} \over  {4 G_N}} \approx   {{ \pi}\over{G_N}} 
\bigg[ P_1P_2 ( Q_1Q_2 -  q^2 )\bigg]^{1\over 2}.
\label{mest}
\end{equation}
One would like to relate the combination of charges
in (\ref{mest})
to the number of `microscopic' string configurations
giving rise to the same black hole solution
at large distances.
In the case of the  fundamental string states of
  toroidally compactified
heterotic string the combination of charges $Q_1Q_2 -  q^2$
would be  related to the number of  the left-moving string oscillation
 modes $N_L$, i.e.
$(Q_1Q_2 -  q^2 )\approx   {1\over 4} \alpha' N_L$
(we  assume that both charges and $N_L$ are large
and set $G_N={1\over 8} \alpha'$, $e^{\Phi'_{\infty}}=1$).
The key  observation is  that in the present case
 the horizon  (throat) region  $r \to 0$
is actually described  by the  $SL(2,R)\times SU(2)$
 WZW-type  model  (\ref{ttr})
with  the level, i.e. the  coefficient in front of
 the action,
$ \kappa=  {4\over \alpha'}  P_1P_2$ (\ref{lev}).
For  large $P_1P_2$, the level $\kappa$ is large  and
the  spectrum of string excitations in this region
should be  approximately  the same
as in the  flat space,
but with  the {\it renormalised} string tension (in the `transverse' part of the action)
\begin{equation}
{1\over \alpha' }\   \to \  {1\over \alpha'_* } =  {P_1P_2\over \alpha' R_1^2}
= {P_1P_2\over \alpha'^2} ,
\label{tens}
\end{equation}
where  we have  set  $R_1=\sqrt{\alpha'}$.
Then
 $Q_1Q_2 -  q^2\approx  {1\over 4} \alpha'_* N_L,$
 or, equivalently,
\begin{equation}
P_1P_2(  Q_1Q_2 -  q^2) \approx {1\over 4} \alpha'^2 N_L .
\label{neww}
\end{equation}
At the same time, the value of the Newton's constant is determined by the
asymptotic $r
\to \infty$ region  and  thus remains unchanged, i.e. $G_N = {\textstyle
{1\over
 8}} \alpha'$.
As a result, the thermodynamic entropy (\ref{mest}) takes
  the form of the  statistical entropy\footnote{The number of   BPS states in the free heterotic string spectrum
with a given  left-moving oscillator number $N_L\gg 1$   is  
$d(N_L)_{N_L\gg 1} \approx  a  N_L^\nu  \exp(2\pi \sqrt {{1\over 6} c_{eff} N_L})$,
where $c_{eff}= D_{crit}-2=24$. In type II theory
$c_{eff}={3\over 2}  (D_{crit}-2)=12$, however,  there are both left- and right-moving 
BPS states in the free string spectrum. 
In contrast  to  the free string case,
the number of BPS oscillation  states   counted 
in our case should be the same 
in the  heterotic and type II theories.  Namely, the relevant 
marginal perturbations 
 are only `left-moving', not `right-moving', i.e. the functions in the sigma-model action
can  depend only on $u$, not on $v$ to preserve the conformal invariance
when both $F$ and $K$  functions are non-trivial.
This is also related to the fact that the
 background 
is `chiral' and has the same amount of space-time supersymmetry 
in both theories, and therefore only `left-moving' perturbations will be  supersymmetric. 
As a result, one should  expect that the entropy  should  be the same in 
heterotic and type II cases, in agreement with the fact that 
the corresponding black hole solution 
and thus  also its thermodynamic  entropy
is the same in two  theories. This suggests that the effective value of the product $c_{eff} N_L$ 
should  be  the same in the two theories.
How this actually 
happens for the $\sigma$-model describing  the horizon region  remains to be 
understood.} \begin{equation}
{\bf S}=  \ln d(N_L)_{N_L \gg 1} \    \approx \  {  4 \pi }  \sqrt { N_L  }  \
{}.
\label{mesti}
\end{equation}
This  argument   generalises the one in \cite{LW}
to the case of an extra  electric parameter $q$
% (the fact that $P_1P_2$
%still factorises  for $q\not=0$ is a  non-trivial consistency  check)
and also explains the `magnetic' renormalisation of the string tension 
\cite{LW}
by direct consideration of the underlying  conformal
model in the horizon (throat) region.

It was  also suggested in  \cite{LW}
that there may exist an interpolating formula  for the entropy which  would be
valid for arbitrary values of charges  and would
reproduce   the stretched  horizon entropy  \cite{sen,peet}
in the limit of vanishing magnetic charges.  The idea  was to
use  the $S$-duality  for the specific example of charge configurations,
obtained from the  generating solution with $q=0$,
 and   to  conjecture  that in general
the $P_1P_2$-factor should be replaced by $P_1P_2 + \alpha' $,
so that the renormalised string tension  should be of  the form
 $ {1\over \alpha'_* } =  {1\over \alpha'} (1+  {1\over \alpha'}  {P_1P_2} ) .$
The  `quantum'  shift by $\alpha'$  can be viewed as
 a  modification of   the purely electric model (where the
area  of the horizon at  $r=0$  is zero)  corresponding to the prescription
of evaluation of the entropy at the stretched horizon at $r=\sqrt{\alpha'}$.
This proposal, however,
does not seem to apply to the  general class of  solutions obtained from the
five-parameter generating solution  with
 $q\not=0$.
In this case one gets a
more  general expression for the  area  of the event horizon  (\ref{mes}),(\ref{genentp})
than the one  assumed  in \cite{LW}.\footnote{According to \cite{LW}
the  level matching condition should
remain essentially
the same as in flat space, i.e. $
\alpha' N_L =4(Q_1Q_2 -q^2+ ...),$
while the magnetic charges  should  enter through  the
 modification of the string
tension  mentioned above.
This  would  imply  that  ${\bf A}\approx 4\pi  \sqrt {N_L}$, \
 $N_L = (1 + {4\over \alpha'}  P_1P_2) [1 + {4\over \alpha'}  (Q_1Q_2 -q^2)]$.
The  general
expression for the  area  (\ref{mes}),(\ref{genentp}) does  not seem to be 
consistent with such  a  factorisation. }
The  general quantum
  formula   for
the entropy,
valid for large as well as small  magnetic charges
and thus interpolating between  the classical general expression
(\ref{genentp}) and  the  one for purely  electric configurations
 evaluated at the stretched horizon
(${\bf A}= 2\pi\sqrt{
2{\vec {\alpha}}^T L{\vec {\alpha}}}$)  \cite{sen,peet},
should   involve  a non-trivial mixture of  electric and magnetic charges.

%%%%%%%%%%%%%%%%%%%%%%%%%%%%%%%%%%%%%%%%%%%%%%%%%%%%%%%%%%
\subsection{Origin of degeneracy: more general `oscillating' solutions}
%%%%%%%%%%%%%%%%%%%%%%%%%%%%%%%%%%%%%%%%%%%%%%%%%%%%%%%%%%%%%%%%%%%%%

The  dyonic black hole  is an  approximate four-dimensional description
of the six-dimensional string soliton
represented  by the conformal model
(\ref{lagr}).
The origin of the `internal degrees of freedom' of the black hole
or a degeneracy   of configurations with  fixed  values of global charges,
which explains the statistical nature of its  entropy,
should  be related to the existence of
%be  in existence of
different six-dimensional string
configurations
which   have the same
%all look the same  from
structure at scales 
larger than  compactification scale.
Such solutions should be represented by
marginal  deformations  of the soliton  theory
which  do not change the values of the  asymptotic
black hole charges.

As in the case of purely electric  BPS-saturated black holes
described by   the five dimensional
fundamental string solutions,
 one should look for more general
conformal models which include (left-moving)
oscillations,  e.g.,  in a  
 compact dimension \cite{CMP,dab}.
Since these  more general solutions  explicitly  depend
on a  compact  internal coordinate,  they can be  represented as
solutions of lower-dimensional theory with {\it massive }
Kaluza-Klein fields having
non-trivial background values.
At scales larger than the
compactification scale these  backgrounds
will  have the same structure as the BPS-saturated black hole,
but the degeneracy
will be lifted once one starts measuring external fields
with resolution  comparable
to the  compactification scale \cite{CMP}.

Like the
oscillating versions of the fundamental string solution
correspond to the  excited  (but still supersymmetric
 BPS-saturated) states of the heterotic string with flat transverse space 
\cite{dh,dab},
 similar    generalisations of the
model (\ref{lagr}),(\ref{latra}) should
 represent  the   BPS-saturated
excited states of the  string soliton with nontrivial transverse (`magnetic')
space.

Remarkably, a class of  supersymmetric
generalisations of the soliton  model (\ref{lagr}),(\ref{latra})
can be obtained  in  the same   way as in the
  fundamental string case,
 by allowing the functions   $K$ and ${\cal A}_i$
in (\ref{lag})  (e.g.  $K$ and  $A$ in (\ref{lagr}))
to depend also on  $u$.\footnote{There  exists, in principle,
  a  possibility of including   a $u$-dependence
also in the functions $G_{ij},B_{ij}, \phi$ in (\ref{lag})
(i.e. in the  functions $f,k$ in  (\ref{latra}))
which define the transverse conformal theory.
In this case  one finds  a non-trivial  second-order
differential
equations  in $u$ \cite{TT} which should be satisfied by  $f(u,x), k(u,x)$.
As a result, only a  special dependence on $u$  may be  allowed in the
`magnetic' part of the model.}
In the case of  the flat transverse  space
this corresponds to the generalised fundamental string solution
with waves traveling along the string as well as fluctuations
in the compact  and non-compact flat  spatial
directions   (see \cite{CMP,dab} and references  therein).

  Starting with the spherically-symmetric
one-center  model
(\ref{FKfk}),(\ref{aaa})
the simplest possibility
is to  add  a $u$-dependent,  but
linear in  $x^s$
 term  in  $K$
(note that the equation for $K$ (\ref{liin})
 does not depend on the functions of
the transverse theory)
and to
replace the constants $Q_1$ and $q$ in  $K$ and $A$
by  arbitrary functions of $u$ (there is no extra constraint
since the term $\partial_u \nabla_i {\cal A}^i $ in   the equation
for $K$ in (\ref{ond}) vanishes in the spherically-symmetric  case)
\begin{equation}
K (u,x) =1 + f_s (u)  x^s  + { Q_1(u) \over r} ,
\ \ \ \  \
 A(u, x) = { q(u) \over r} \cdot { r +  {1\over 2} (P_1 + P_2) \over r + P_1} .
\label{perr}
\end{equation}
Here $  Q_1(u) =  Q_1  + \tilde Q_1(u),   \
q(u)= q+ \tilde
 q(u)$.\footnote{The asymptotic flatness of the background can be
restored by a coordinate
transformation  ($x^s \to x^s - \tilde f^s (u), \  \partial^2_u \tilde f^s 
\sim f^s,$
etc.) as in \cite{CMP,dab}.
Note also that in general the $u$-dependent part of $K$ can be traded for 
the ${\cal A }_i$-perturbation in (\ref{lag})
by making a redefinition of $v$ \cite{TH}.}

For simplicity,   let us   ignore oscillations
in the non-compact dimensions, i.e. $f_s=0$.
In the   fundamental string  case ($P_1=P_2=0$)
 one  expects  the `matching condition'\footnote{It can 
be
imposed either by
requiring  that the $r=0$ singularity of the higher-dimensional
background
should be null \cite{CMP}
 or by  using  string source considerations \cite{dab} and $T$-duality (see
also the next footnote).}
\begin{equation}
 Q_1 (u) Q_2  - q^2 (u)  = 0,
\label{metc}
\end{equation}
which, after averaging in the compact coordinate $u$,
can  be put in the form
of the (classical)   level matching condition  for  the
 elementary string states\footnote{This relation should already hold
 in the bosonic string case.
In fact, $Q_1$ plays the role of the momentum along the string.
Since $Q_1$ and $Q_2$ are interchanged by $T$-duality in $u=y_2$-direction,
$Q_2$  should be  an analogue
of the winding number. Then  their product
should be proportional to the difference  of  the left-moving ($N_L$) 
and right-moving ($N_R$)
oscillation numbers. In heterotic string
case  there are no  classical  oscillations  in the right-moving sector, i.e.
 $N_R=0$.}
\begin{equation}
 {4 \over \alpha'}(Q_1Q_2  - q^2)  =    N_L     ,
\ \ \ \   N_L   \equiv  {4 \over \alpha'} < \tilde q^2 (u) > .
\label{speh}
\end{equation}
Analogous  level matching condition  should exist
for excited states of the soliton theory
with non-vanishing $P_{1,2}$.\footnote{Although
the  higher-dimensional
 soliton background is non-singular at $r=0$  if $Q_1 Q_2P_1P_2\not=0$
 it may still be possible to derive
such a level matching condition from some
geometrical considerations,
  cf.  \cite{LW}.}
Since now the horizon region is described by
a well-defined conformal theory,
the constraint should
be  just  the level matching condition
for  the corresponding  states of
the generalised   WZW-type theory   (\ref{ttr}).
For example,
replacing $q$ in  (\ref{thro}),(\ref{not})
by a periodic function of $u=y_2$,  $
q \to q + \tilde q(y_2) $
 corresponds to adding to (\ref{thro})
the  perturbation (cf. (\ref{ttr}),(\ref{not}))
\begin{equation}
2  \tilde q(y_2)
  \partial \tilde y_2 [{\bar\partial} y_1 + P_1 (1-\cos \theta) {\bar\partial} 
\varphi]= 2 P_1  
  \tilde q(y_2)
  \partial \tilde y_2  \bar J_3.
\label{yyyy}
\end{equation}
It   is marginal  for any  $\tilde q(y_2)$ 
because of the presence of the  $e^{-z} \partial y_2 {\bar\partial} t$-term
in  (\ref{ttr}).
By analogy with the fundamental string case
we expect that  for large level $\kappa$ (\ref{lev})
the corresponding state in the  soliton
string spectrum should  also satisfy
\begin{equation}
\kappa (Q_1Q_2  - q^2) =
 {4 \over \alpha'} P_1P_2 (Q_1Q_2  - q^2)   \sim  < \tilde q^2 (u) >  \sim N_L.
\label{fff}
\end{equation}
The  main  difference compared to 
 (\ref{speh})
 is  again the renormalisation of the
string tension by $P_1P_2$,
in agreement with the suggestion in \cite{LW}.
The relation  (\ref{fff}) provides an interpretation  of 
(part of)  the  degeneracy  $N_L$  in  (\ref{neww}) in terms of 
 (classical)
 oscillations of the underlying soliton
in the internal $y_2$-direction.
Further study of  marginal BPS-saturated 
perturbations
of the soliton model
is important for  making
the statistical interpretation of the black hole entropy
(\ref{mest}) more quantitative.

%\vskip2.mm
%\vskip2.mm
%%%%%%%%%%%%%%
\acknowledgments
%{\bf Acknowledgements}
%%%%%%%%%%%%%%%%%%%%%%%%%%%%%%%%%%%%%%%%%%%%%%%%%

We would like to thank K. Chan, C. Hull, F. Larsen, A. Peet, A. Sen, K. 
Sfetsos,  F. Wilczek, D. Youm, and
especially
 E. Witten  for useful discussions.
A.A.T. is grateful to the HEP group of Physics Department of Princeton
University and Theory Division of CERN for hospitality while this work was in
progress.
The work of M.C.  is supported  by  the Institute for Advanced Study funds and
J. Seward
Johnson foundation,  U.S. Department of Energy Grant No.
DOE-EY-76-02-3071, and the National Science
Foundation Career Advancement Award PHY95-12732.
The work of A.A.T.   is supported  by
 PPARC  and
 ECC grant SC1*-CT92-0789.
We also acknowledge  the support of
the NATO collaborative research grant CGR 940870.

%%%%%%%%%%%%%%%%%%%%%%%%%%%%%%%%%%%%%%%%%%%%%%%%%%


\vskip2.mm
\begin{references}

%\begin{thebibliography}{19}

\bibitem{gib} {G.W.  Gibbons,  Nucl. Phys. {\bf B207}  (1982) 337;
G.W.  Gibbons and K. Maeda, Nucl. Phys. {\bf  B298} (1988) 741;
 D. Garfinkle, G.T.  Horowitz and A. Strominger, Phys. Rev. {\bf D43} (1991)
3140, Erratum, {\it ibid.} {\bf D45} (1992) 3888. }

\bibitem{HOR}{G.T.  Horowitz, in {\it
Proceedings of the 1991 Trieste Spring School on String Theory
and Quantum Gravity} (World Scientific, Singapore 1993), hep-th/9210119.}



\bibitem{TS}{A.A. Tseytlin, Class. Quant. Grav. {\bf 12}  (1995) 2365, 
hep-th/9505052.}


\bibitem{seen} {A. Sen, Nucl. Phys. {\bf B440} (1995) 421, hep-th/9504027. }


\bibitem{HRT}{ G.T. Horowitz and A.A. Tseytlin, {Phys. Rev. Lett.}
 {\bf 73 } (1994) 3351,
hep-th/940840.}

\bibitem{TH}{ G.T. Horowitz and A.A. Tseytlin, Phys. Rev. {\bf D51} (1995)
2896, hep-th/9409021.}

\bibitem{BE}{ K. Behrndt, Nucl. Phys.  {\bf B455} (1995) 188, hep-th/950610.}

\bibitem{CMP}{ C.G. Callan, J.M.  Maldacena  and A.W. Peet,
%``Extremal black holes as fundamental strings",
PUPT-1565,  hep-th/9510134.  }

\bibitem{dab}{  A. Dabholkar, J.P. Gauntlett, J.A. Harvey and D. Waldram,
%``Strings as solitons and black holes as strings",
 CALT-68-2028, hep-th/9511053.   }

\bibitem{dh}{A. Dabholkar and J.A.  Harvey, {Phys. Rev. Lett.} {\bf 63} (1989) 
478.}

\bibitem{duh}{A. Dabholkar, G.W.   Gibbons, J.A.   Harvey  and F. Ruiz-Ruiz,
Nucl. Phys. {\bf B340} (1990) 33. }

\bibitem{ME}{ A.A. Tseytlin, 
 Phys. Lett. {\bf B}363 (1995) 223,  hep-th/9509050. }

\bibitem{MME} { A.A. Tseytlin, Phys. Lett. {\bf B251} (1990) 530.}

\bibitem{sen}{A. Sen,  Mod. Phys. Lett.  {\bf A10} (1995) 2081,
 hep-th/9504147. }

\bibitem{suss}{L. Susskind, RU-93-44, hep-th/9309145;
L. Susskind and J. Uglum, Phys. Rev. {\bf D50} (1994) 2700, hep-th/9401070;
J. Russo and L. Susskind,  Nucl. Phys. {\bf B437} (1995) 611, hep-th/9405117.}

\bibitem{peet}{A.W.  Peet,  Nucl. Phys. {\bf B456} (1995) 732,
 hep-th/9506200.}

\bibitem{nel}{W.  Nelson, Phys. Rev. {\bf D49} (1994) 5302, hep-th/9312058. }

\bibitem{kalor}{R. Kallosh  and  T. Ortin, Phys. Rev.  {\bf D50} (1994)
7123, hep-th/9409060.}

\bibitem{CHSR} {C.G. Callan, J.A. Harvey and A. Strominger, in {\it
Proceedings of the 1991 Trieste Spring School on String Theory and
Quantum Gravity}, J.A. Harvey {\it et al. eds.} (World Scientific,
Singapore
1992).}

\bibitem{dukh}{ M.J. Duff, R.R. Khuri and J.X. Lu,  Phys. Rep. {\bf 259}
(1995) 213, hep-th/9412184. }


\bibitem{chs}{C.G. Callan, J.A. Harvey and A. Strominger,
Nucl. Phys.  {\bf B359 } (1991)  611.}


\bibitem{CY}{M. Cveti\v c and D. Youm, Phys. Rev. {\bf D53} (1996) 584, 
hep-th/9507090.}

\bibitem{CYI}{M. Cveti\v c and D. Youm,  Phys. Lett. {\bf B359} (1995) 87,
hep-th/9507160.}

\bibitem{CYII}{M. Cveti\v c and D. Youm, UPR-0675-T, to be published in the
{\it Proceedings of  STRINGS 95: Future Perspectives in String Theory}
(World Scientific,  1996), hep-th/9508058.}


\bibitem{US}{M. Cveti\v c and  A.A.  Tseytlin,
 Phys. Lett. {\bf B366} (1996) 95, hep-th/9510097.
}

\bibitem{LW}{ F. Larsen  and F. Wilczek,
%``Internal structure of black holes",
PUPT-1576,  hep-th/9511064.    }



%\bibitem{TSH}{ G.T. Horowitz and A.A. Tseytlin, Phys. Rev. {\bf D50} (1995)
%5204, hep-th/9406067.}



\bibitem{TT}{ A.A. Tseytlin, Nucl. Phys. {\bf B390} (1993) 153, 
hep-th/9209023. }


\bibitem{garf}{D. Garfinkle,  Phys. Rev.  {\bf D46} (1992) 4286.}

\bibitem{senn} {A. Sen, Nucl. Phys. {\bf B450} (1995) 103,
hep-th/9504027.}

 \bibitem{HAS}{J.A.  Harvey  and A. Strominger, Nucl. Phys. {\bf B449 }
(1995) 535, hep-th/9504047. }

\bibitem{DL}{M.J.  Duff and J.X. Lu,  Nucl. Phys. {\bf B354}  (1991) 141.}

 \bibitem{duf}{M.J.  Duff, S. Ferrara, R.R. Khuri and J. Rahmfeld, Phys. Lett. 
{\bf
B356}
(1995) 479, hep-th/9506057.}


\bibitem{HTI}{C.M.  Hull and P.K. Townsend, Nucl. Phys. {\bf  B438} (1995)
109, hep-th/9410167.}

\bibitem{WITTENII} {E. Witten, Nucl. Phys. {\bf B443} (1995) 85,
hep-th/9503124.
}

\bibitem{LS}{D.A. Lowe and A. Strominger,  Phys. Rev. Lett. {\bf  73}
 (1994) 1468, hep-th/9403186.}
 
\bibitem{gps}{S. Giddings, J.  Polchinski and A.
 Strominger,  Phys. Rev. {\bf D48} (1993) 5784, hep-th/9305083.}

\bibitem{MS}{J. Maharana and J.H.   Schwarz, Nucl. Phys. {\bf B390} (1993) 3,
 hep-th/9207016.  }

\bibitem{CYIII}{M. Cveti\v c and D. Youm, IASSNS-HEP-95-107, hep-th/9512127.}

\bibitem{GHT}{G.W. Gibbons, G.T. Horowitz and  P.K. Townsend,      Class. Quant. Grav. {\bf 12}  (1995) 297.}

\bibitem{DLR}{M.J. Duff, J.T. Liu and J. Rahmfeld, CTP-TAMU-27-95,
 hep-th/9508094.}


\bibitem{SEN1}{A. Sen, Int. J. Mod. Phys. {\bf A9} (1994) 3707,
hep-th/9402002.}

\bibitem{WITTENIII}{E. Witten, Phys. Lett. {\bf B86} (1979) 283.}


\bibitem{CYIV}{M. Cveti\v c and D. Youm,  Phys. Rev. {\bf D52  } (1995)
2144, hep-th/9502099 and  Phys. Rev. Lett. {\bf 75 }(1995) 4165 ,
 hep-th/9503082.}



%\end{thebibliography}



\end{references}
\end{document}